\documentclass[fleqn,usenatbib]{mnras}

\usepackage{newtxtext,newtxmath}
\usepackage{CJKutf8}
\usepackage{xcolor}
\usepackage[T1]{fontenc}
\usepackage{graphicx}
\usepackage{amsmath}

\newcommand{\orcid}[2]{\href{http://orcid.org/#2}{#1}}
\newcommand{\kms}{\,km\,s$^{-1}$}
\newcommand{\mstar}{{\mbox{$M_{\rm star}$}}}
\newcommand{\msun}{{\mbox{$M_{\odot}$}}}
\newcommand{\lya}{\mbox{${\rm Ly}\alpha$}}

\title[Ionized Gas in NGC\,3511 and NGC\,3513]{Spatially Resolved Properties of Extraplanar Diffuse Ionized Gas in NGC\,3511 and NGC\,3513}

\author[Zhu, Boettcher, and Chen]{
\orcid{Hanjue Zhu~\begin{CJK*}{UTF8}{gkai}(朱涵珏)\end{CJK*}}{0000-0003-0861-0922},$^{1}$\thanks{E-mail: hanjuezhu@uchicago.edu}
\orcid{Erin Boettcher}{0000-0003-3244-0409},$^{2,3,4}$
and \orcid{Hsiao-Wen Chen~\begin{CJK*}{UTF8}{bkai}(陳曉雯)\end{CJK*}}{0000-0001-8813-4182}$^{1}$
\\
$^{1}$Department of Astronomy \& Astrophysics, The University of Chicago, Chicago, IL 60637, USA\\
$^{2}$Department of Astronomy, University of Maryland, College Park, MD 20742, USA\\
$^{3}$X-ray Astrophysics Laboratory, NASA/GSFC, Greenbelt, MD 20771, USA\\
$^{4}$Center for Research and Exploration in Space Science and Technology, NASA/GSFC, Greenbelt, MD 20771, USA
}

\date{Submitted to MNRAS}
\pubyear{2024}

\begin{document}
\label{firstpage}
\pagerange{\pageref{firstpage}--\pageref{lastpage}}
\maketitle

\begin{abstract}
Gaseous, disk-halo interfaces are shaped by processes that are critical to galaxy evolution, including gas accretion and outflows. Extraplanar diffuse ionized gas (eDIG) layers are characterized by scale heights that largely exceed those predicted by their temperature, suggesting the presence of turbulent energy injection from star formation feedback. However, the origin of this large scale height remains uncertain. To explore the connection between eDIG and star-forming disks, we present a spatially-resolved case study of a nearby pair of sub-$L_*$, intermediately inclined disk galaxies NGC\,3511/3513. We decompose optical nebular lines observed using long-slit spectroscopy into narrow and broad velocity components. In NGC\,3511, the broad component has three distinctive characteristics in comparison to the narrow component: (1) significantly higher velocity dispersions (a median $\langle\sigma\rangle_{\text{Broad}} = 24$ \kms compared to $\langle\sigma\rangle_{\text{Narrow}} = 13$ \kms), (2) elevated [\ion{N}{II}]$\lambda$6583/H$\alpha$ and [\ion{S}{II}]$\lambda$6716/H$\alpha$ line ratios, and (3) a rotational velocity lag. These characteristics support the origin of the broad component in an extraplanar, gaseous disk. In NGC\,3513, the broad component reveals disk-halo circulation via localized outflows at radius $\lesssim 1$ kpc. For NGC\,3511, we test a vertical hydrostatic equilibrium model with pressure support supplied by thermal and turbulent motions. Under this assumption, the eDIG velocity dispersion corresponds to a scale height $h_{z} \gtrsim 0.2 - 0.4$ kpc at $R = 3 - 5$ kpc, a factor of a few above the thermal scale height ($h_{z} \lesssim 0.1$ kpc). This highlights the importance of turbulent motions to the vertical structure of the gaseous, disk-halo interface.
\end{abstract}


\section{Introduction}
The exchange of gas, metals, and energy through a boundary layer where the galactic disk transitions into the galactic halo, i.e. the gaseous disk-halo interface, drives the evolution of disk galaxies. Gas accretion is needed to replenish galaxies' gas supplies and sustain star formation, while star-formation feedback expels chemically-enriched gas into the surrounding circumgalactic medium (CGM) and intergalactic medium (IGM).  Together, these processes determine the star-formation histories of galaxies and affect their chemical abundances. Since gas passes through the disk-halo interface in all of the accretion and feedback processes, studying the density distribution and kinematics of the disk-halo interface gives important information about these key processes driving galaxy evolution.

Across the disk–halo interface, there is a transition in the phase, density, and kinematics of the gas. Early observations revealed the presence of a diffuse layer of warm ionized gas in our own Milky Way, known as the Reynolds Layer \citep{Hoyle1963,Reynolds1973}. \citet{Reynolds1989} found that this ionized gas has unexpectedly large scale height ($h_{z} \approx 1$ kpc) compared to its thermal scale height ($h_{z} \approx 0.1$ kpc at $R = 3$ kpc), calculated assuming a hydrostatic equilibrium supported by only thermal motions ($T \sim 10^{4}$ K, $\sigma_{\text{th}} \sim 10$ km s$^{-1}$). In many nearby galaxies, analogous distributions of extraplanar diffuse ionized gas (eDIG) have been found \citep[e.g.,][]{Dettmar1990, Rand1990, Lehnert1995, Veilleux1995, Hoopes1999, Miller2003, Miller2003b, Rossa2003, Rossa2003b, Haffner2009, Jones2017, Levy2019}. These diffuse gas layers are vertically extended and photoionized with a typical temperature of $T \sim 10^{4}$ K and an inferred mid-plane mean electron density of $\langle n_{e,0} \rangle \sim 0.1$ cm$^{-3}$.

\begin{table*}
	\centering
	\caption{Physical properties of the galaxy pair NGC\,3511 and NGC\,3513.}
	\label{tab:galaxy_properties}
	\begin{tabular}{lccccccc}
		\hline
		 & & Distance$^a$ & $r_{25}^b$ & $i$ & SFR\,$^c$ &  &  \\
		Galaxy & Redshift & (Mpc) & (arcmin) & (deg) & ($M_\odot$/yr) & $\log\,($\mstar$/M_\odot$)$^d$ & log(Specific SFR/yr$^{-1}$) \\
		\hline
		NGC\,3511 & 0.00370 & $13.94 \pm 2.10$ & 6.0 & 74.5 & 0.8 & $10.0 \pm 0.1$ & $-10.1$ \\
		NGC\,3513 & 0.00398 & $13.94 \pm 2.10$ & 2.8 & 39.0 & 0.2 & $9.2 \pm 0.1$ & $-9.8$ \\
		\hline	
	\end{tabular}
	\parbox{\textwidth}{\small $^\mathrm{a}$ Distance determined with the local flow model of \citet{Kourkchi2020}, based on Cosmicflows measurements of \citet{Tully2016}.}\\
	\parbox{\textwidth}{\small $^\mathrm{b}$ Disk diameter corresponding to $\mu_B=25$ mag\,arcsec$^{-2}$ isophot \citep{Makarov2014}.}\\
	\parbox{\textwidth}{\small $^\mathrm{c}$ SFR based on FUV and MIR observations with an uncertainty of 0.2 dex \citep{Leroy2019, Leroy2021}.}\\
	\parbox{\textwidth}{\small $^\mathrm{d}$ \mstar\ reported by \citet{Leroy2019} and \citet{Leroy2021}.}
\end{table*}

The dynamic state of the eDIG layers and its implications are not yet understood. Proposed mechanisms for the eDIG origin include (a) internally, the star-formation driven disk-halo flow (e.g., the superbubble breakout model \citep{MacLow1988}, the galactic chimney model \citep{Norman1989}, the galactic fountain model \citep{Shapiro1976}, and the galactic wind model \citep[][and references therein]{Veilleux2005}), and (b) externally, gas accretion from the IGM \citep[e.g.,][and references therein]{Binney2005,Putman2017}, or the hot halo \citep{Marasco2012,Li2023}. Numerous models exist to explain the vertical structure and kinematics of the eDIG layers. A simple first step is to determine the ingredients needed to generate sufficient vertical pressure to stably support eDIG layers under a dynamic equilibrium. Hydrostatic equilibrium models \citep[e.g.,][]{Boulares1990,Barnabe2006,Marinacci2010} have attempted to reproduce the observed kinematics of extraplanar gas in edge-on galaxies. While these models qualitatively reproduce the rotational velocity gradients in nearby galaxies, they require a gas temperature above that of the warm neutral and ionized gas. In a series of papers
\citep{Boettcher2016,Boettcher2017,Boettcher2019}, individual case studies were conducted on nearby galaxies to investigate whether or not turbulent, magnetic field, and cosmic-ray pressure gradients, in addition to the thermal pressure gradient, are sufficient to support the eDIG layers at the observed scale heights. However, constraints on the \textit{vertical} kinematics of the eDIG have been limited to date \citep[e.g.,][]{Fraternali2004, Boettcher2017,Li2021}, hindering our understanding of eDIG dynamics perpendicular to their host galaxies. In this paper, we apply such analysis to a nearby galaxy pair NGC\,3511 and NGC\,3513.

NGC\,3511 is a nearby disk galaxy ($z\approx 0.0037$; \citealt{Koribalski2004}) that is highly inclined with an inclination angle of $i = 74.5^{\circ}$ (\citealt{Makarov2014}). It is close to the companion galaxy NGC\,3513, with a projected separation of $d_{\rm proj} = 50$ kpc. NGC\,3513 is a nearly face-on ($i = 39.0^{\circ}$) SBc spiral at $z\approx 0.00398$ \citep{Kaldare2003}. NGC\,3511 and NGC\,3513 have star formation rates (SFRs) derived jointly from the far-ultraviolet (FUV) and mid-infrared (MIR) of SFR $= 0.8$ $M_\odot \rm \,yr^{-1}$ and SFR $ = 0.2$ $M_\odot \rm \,yr^{-1}$, respectively \citep{Leroy2019, Leroy2021}. NGC\,3511 and NGC\,3513 have stellar masses of $\mstar = 10^{10.0} M_\odot$ and $\mstar = 10^{9.2} M_\odot$, respectively, with a mass ratio of $\approx 6:1$ \citep{Leroy2019, Leroy2021}. The physical properties of the galaxy pair are summarized in Table~\ref{tab:galaxy_properties}. \citet{Rossa2003} identified thresholds in two far-IR (FIR) SFR indicators, the FIR surface brightness and the FIR color at $60$ and $100$ $\mu$m, above which eDIG is detected almost ubiquitously in H$\alpha$ imaging surveys of edge-on galaxies. NGC\,3511 and NGC\,3513 are selected to fall above the thresholds where eDIG is expected to be detected.\footnote{We adopt $S_{60}$ and $S_{100}$, the $60$ and $100$ $\mu$m flux densities, as reported by \citet{Surace2004}. NGC\,3511 has $\log S_{60} / S_{100} =  -0.41$ and $\log L_{\rm FIR} / D_{25}^2 = 0.70$; NGC\,3513 has $\log S_{60} / S_{100} = -0.37$ and $\log L_{\rm FIR} / D_{25}^2 = 0.94$, where the latter quantity inside the log has units of $10^{40}$ erg s$^{-1}$ kpc$^{-2}$.}

\begin{figure}
    \centering
    \includegraphics[width=0.99\columnwidth]{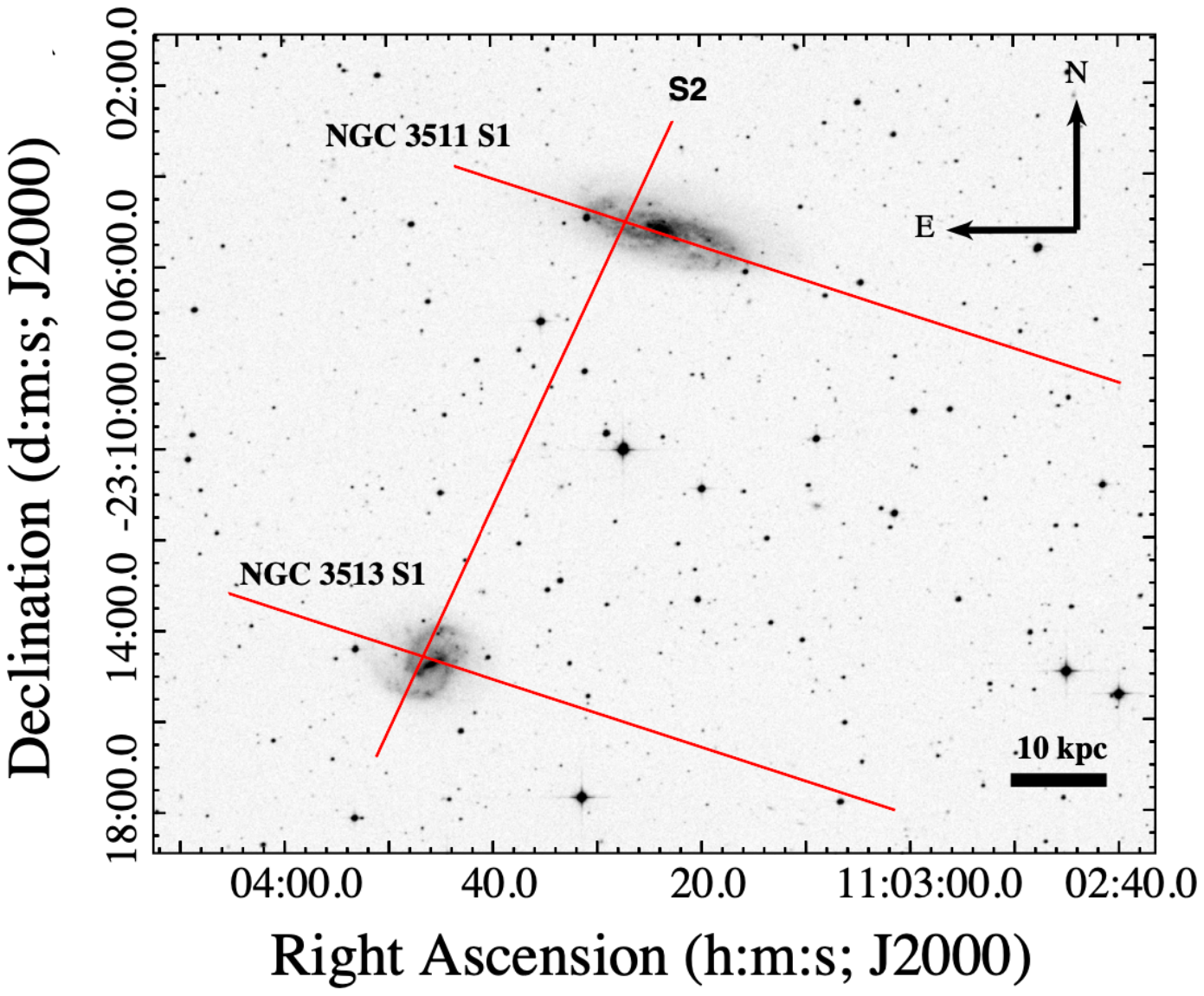} 
\caption{The three longslits overlaid on a red image of NGC\,3511 and NGC\,3513 from the Digitized Sky Survey (\url{https://archive.stsci.edu/cgi-bin/dss_form}). Red data are available from all three slits; blue data are available for NGC\,3511 S1 only. The QSO sightline (outside the field of view) is located $68$ kpc southwest of NGC\,3513. The sightline is near the minor axis of NGC\,3511, at a projected distance of $\rho = 112$ kpc from NGC\,3511.}
\label{fig:slit_pos}
\end{figure}

This galaxy pair is associated with a \lya\ absorber ($\log\,N(\ion{H}{I})/{\rm cm}^{-2} = 14.8$) identified along the sightline toward QSO PMN J1103$-$2329 \citep{Keeney2017}. The QSO sightline is outside the field of view of Figure~\ref{fig:slit_pos} and is located $68$ kpc southwest of NGC\,3513. The sightline is near the minor axis of NGC\,3511, at a projected distance of $\rho = 112$ kpc from NGC\,3511.). The association of the detected absorber with one galaxy or the other is ambiguous. \citet{Keeney2017} showed that the \lya\ absorber can be decomposed into two components at $cz_{\text{abs}} = 1113$ \kms and 1194 \kms, which coincide with the systemic velocities of NGC\,3511 and NGC\,3513 at $cz_{\text{sys}} = 1114$ km s$^{-1}$ and $1194$ km s$^{-1}$, respectively. Additionally, their photoionization analysis indicates that the gas is highly enriched to twice solar, which is surprisingly high considering the large distances from the nearest galaxies. Our studies of the extraplanar gas layers in this galaxy pair may provide additional insight into the circulation of mass and metals between the galaxies and their gaseous environment.

This paper is organized as follows. In Section~\ref{sec:obs}, we discuss the data acquisition and reduction procedures. In Section~\ref{sec:methodology}, we detail our spectral fitting strategies and our identification of the eDIG components in optical emission lines. In Section~\ref{sec:results}, we show the observational results, which strongly support that the eDIG components are extraplanar in origin. In Section~\ref{sec:model}, we build a dynamic equilibrium model of the eDIG layer in NGC\,3511 and show that turbulent motions provide significant support to the eDIG layer. In Section~\ref{sec:discussion}, we discuss the impact of star formation activities on the dynamics and morphology of the eDIG and explore the possibility of associating the galaxy pair with the nearby absorber. Throughout the paper, we adopt a $\Lambda$ cosmology with $\Omega_{\text{M}} = 0.3$, $\Omega_{\Lambda} = 0.7$, and $H_{0} = 70$ km\,s$^{-1}$ Mpc$^{-1}$.

\section{Observations \& DATA REDUCTION} \label{sec:obs}

To characterize the dynamic state of diffuse gas in galaxies NGC\,3511/3513, we have obtained spatially-resolved high-resolution optical spectra along the disks of these galaxies.  In this section, we summarize the observations and data reduction.

\begin{table}
	\centering
	\caption{Summary of observations.}
	\label{tab:obs_details}
	\begin{tabular}{lcccr}
		\hline
		& RA$^a$ & Dec$^a$ & P.A.$^b$ & \multicolumn{1}{l}{$t_{\rm{exp}}$}\\
		Target + Slit Name& (J2000) & (J2000) & (deg) & \multicolumn{1}{c}{(sec)}\\
		\hline
		NGC\,3511 S1 & 11:03:27.13 & $-$23:05:02.18 & 72 & 3$\times$1800\\
		NGC\,3511/3513 S2 & 11:03:46.65 & $-$23:14:35.17 & 72 & 3$\times$1800\\
		NGC\,3513 S1 & 11:03:46.65 & $-$23:14:35.17 & 335 & 3$\times$1800\\
		\hline	
	\multicolumn{5}{l}{$^\mathrm{a}$ The RA and Dec at the center of the slit.} \\
	\multicolumn{5}{l}{$^\mathrm{b}$ The position angle measured from North to East.} \\
	\end{tabular}
\end{table}

\subsection{Observations}
\label{sec:data_collection}
 Optical spectroscopy of NGC\,3511 and NGC\,3513 was carried out on 2019 May 01 - 02 and on 2020 March 16 using a $0.7''$ long slit and the f/4 camera in the Inamori-Magellan Areal Camera \& Spectrograph (IMACS) multi-object imaging spectrograph \citep{Dressler2011} on the Magellan Baade Telescope.  IMACS f/4 covers a field of view of $15'$ in diameter with a spatial plate scale of $0.111''$/pixel. The typical seeing was $0.6 - 0.7''$ in 2019 and $0.4 - 0.6''$ in 2020.  In May 2019, the observations were performed using the $1200 \,\ell\,{\rm mm}^{-1}$ grating with a blaze angle of $26.7^{\circ}$. The tilt angle was chosen to match the blaze angle, which yielded  a dispersion of 0.194 \AA/pixel and a resolution of $R\sim7140$ (or equivalently $\Delta\,v_{\rm FWHM} \approx$ 42 \kms) at around 6600 \AA. In March 2020, the $1200 \,\ell\,{\rm mm}^{-1}$ grating with a blaze angle of $17.5^{\circ}$ was used at a tilt angle of $16.7^{\circ}$, which yielded a dispersion of 0.194 \AA/pixel and $R \sim5770$ (or equivalently $\Delta\,v_{\rm FWHM} \approx$ 52 \kms) at around 5000 \AA. 
 
 With the grating setup in the May 2019 observing run, the spectroscopic data cover emission features due to [\ion{N}{II}] $\lambda$6548, 6583, H$\alpha$, and [\ion{S}{II}] $\lambda$6716, 6731 emission lines from the galaxies.  The data obtained in March 2020 provide additional spectral coverage for H$\beta$ and [\ion{O}{III}] $\lambda$4959, 5007 emission lines. Hereafter, we refer to the May 2019 data as the ``red data'' and the March 2020 data as the ``blue data''.  In May 2019, the observations were carried out in three slit positions, two of which were aligned along the major axis of each of the two galaxies to maximize the information collected at different galactocentric radii (S1). A third slit position was employed to cover both galaxies at once.  This third slit position is also approximately perpendicular to the first two slit positions. In March 2020, we placed the slit along the major axis of NGC\,3511. Note that although we intended to place the slit masks at the same locations in 2019 and 2020, there is a spatial offset of about one slit width ($0.7''$). The locations of S1 and S2 relative to the two galaxies are illustrated in Figure~\ref{fig:slit_pos}. A brief journal of the observations is given in Table~\ref{tab:obs_details}.

\subsection{Data Reduction}
We used custom software developed exclusively for this project to reduce the IMACS data. The two-dimensional spectra were bias subtracted using the overscan region and corrected for flat-field variations using a spectral flat obtained before each set of science exposures. Sky subtraction was performed for each exposure using a model sky spectrum constructed from sky regions selected to exclude contamination from the galaxy. Individual sky-subtracted frames were then median combined to form a final stack. Throughput correction and flux calibration were then performed using a standard star observed on the same night (LTT3864 in May 2019 and EG274 in March 2020). In addition, wavelength solutions were obtained using the HeArNe comparison lamp spectrum taken before each set of science exposures.  These arc lines also provided an empirical measure of the spectral resolution of the adopted slit-grating combination. 

We extracted one-dimensional (1D) spectra using an aperture of 3 spatial pixels (0.33'', or 0.02 kpc) along the cross-dispersion direction. This aperture size is both sufficiently small to ensure that we do not introduce artificial eDIG signals due to velocity shear across the aperture and moderately large to boost the signal-to-noise ratio (S/N) of the 1D spectra. For each spectrum, we estimate the flux uncertainty assuming that S/N scales according to Poisson statistics (uncertainty=$\sqrt{N_\gamma}$, where $N_\gamma$ is the flux count in units of electrons). Finally, we perform continuum subtraction by masking the emission lines and adopting a local, linear model for the continuum adjacent to the emission lines. For NGC\,3511, we fit a total number of 1124 spectra (844 for S1 and 280 for S2) and 609 for NGC\,3513 (262 for S1 and 347 for S2).

\section{Methodology for line profile analysis} \label{sec:methodology}

The optical spectra of NGC\,3511 and NGC\,3513 exhibit prominent nebular lines typical of \ion{H}{II} regions.  By measuring the line centroid and line width, we are able to delineate contributions from local star-forming regions and from the eDIG layer.  In this section, we describe the steps we take to map the gas motion at different locations along the disks of these galaxies and to estimate the associated uncertainties.

\subsection{Emission-line measurements}
From previous studies, we expect that the velocity dispersion in the diffuse gas may be higher than in the \ion{H}{II} regions \citep[e.g.,][]{Veilleux1995,Heald06b,Heald07,Boettcher2016,Boettcher2017, Boettcher2019}. This motivates us to model the emission lines using a two-component Gaussian model with a narrow component representing the planar emission and a broad component for the extraplanar gas. 

In addition, the components sharing common origins are expected to exhibit similar velocity profiles.  Therefore, the broad and narrow components of different nebular lines are assumed to share the same velocity centroid and dispersion in the model profiles. The results of the emission line measurements include velocity centroid, line width, total line flux, continuum level, and line ratios. Because the instrumental resolution depends only weakly on wavelength, we correct the observed line width for the instrumental broadening using the same instrumental line width across all nebular lines.

\subsection{Uncertainty estimates} 

To obtain robust uncertainties for the best-fit parameters of our multi-Gaussian fits, we employ a Markov Chain Monte Carlo (MCMC) method. We use an open-source Python implementation of Affine Invariant \citep{GoodmanandWeare} MCMC Ensemble sampler \textit{emcee} \citep{emcee} to sample the posterior distribution. We use the median of the posterior for the best-fit value, with the uncertainty quantified by the 84th and 16th quantiles, representing the upper and lower bounds, respectively. \textit{emcee} uses the integrated autocorrelation time (IAT) to assess convergence. IAT is an estimate of the number of iterations for an independent sample to be drawn in a Markov chain. That is, the number of independent samples is of order the number of samples divided by the IAT, which we denote as $\tau$ ($N_{\rm independent} \sim N/\tau$). In our line fitting, we declare convergence when $\tau$ is smaller than $N/50$, as suggested by the \textit{emcee} documentation. We run our ensemble sampler with 100 walkers and 60,000 steps for every spectrum we fit. Most of the chains are converged by the IAT criterion. For the several cases that do not, it is usually because there are some unphysical local minima in the posterior. Also, note that we assume a Gaussian PDF to construct the likelihood function.

\begin{figure*}
    \centering
    \includegraphics[width=\textwidth]{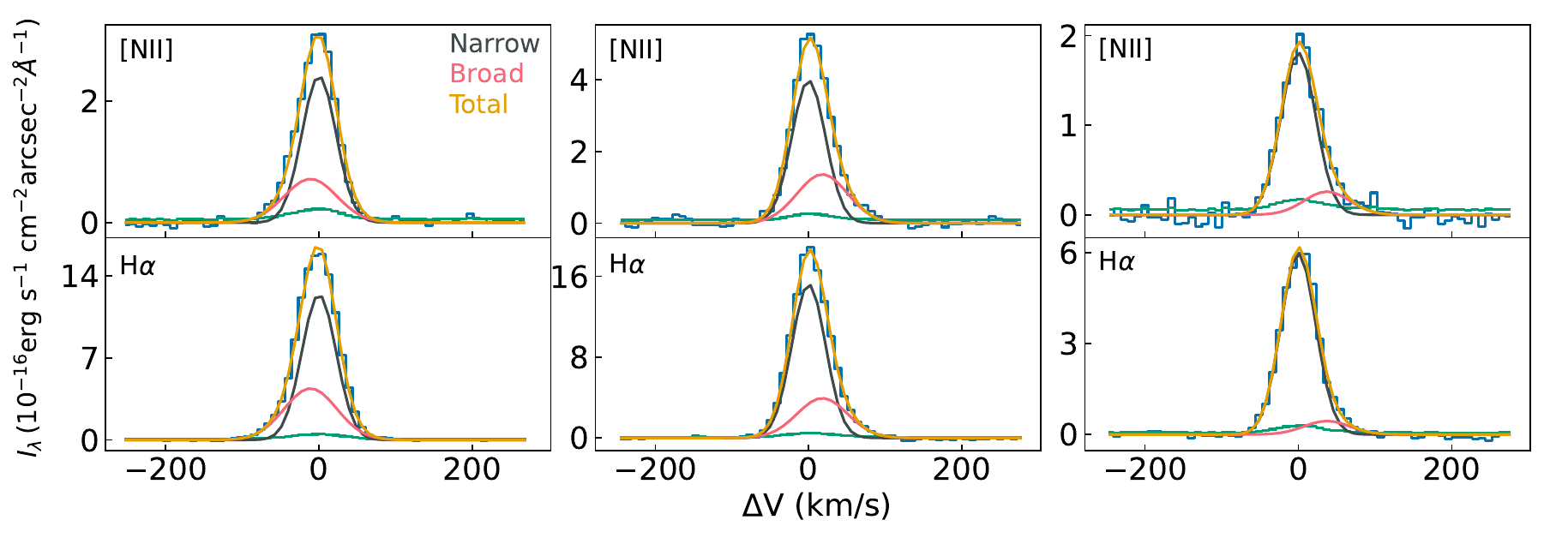}
\caption{Example [\ion{N}{II}] $\lambda$6583 and H$\alpha$ emission lines obtained from three regions along S1 of NGC\,3511. The data and the uncertainties are shown in blue and green, respectively. We decompose each spectrum into a narrow (grey) and a broad (pink) component; the combined model is shown in yellow. We fit [\ion{N}{II}] $\lambda$6583, H$\alpha$, and [\ion{S}{II}] $\lambda$6716, 6731 simultaneously but we only show two lines here for clarity. These examples are chosen to represent the diversity in the emission-line intensities and ratios and the velocity structure observed in the data.}\label{fig:example_spec}
\end{figure*}

Using \textit{emcee}, for the red data, we fit the [\ion{N}{II}] $\lambda$6583, H$\alpha$, and [\ion{S}{II}] $\lambda$$\lambda$6716, 6731 emission lines at once. We do not include the [\ion{N}{II}] $\lambda$6548 lines due to generally low S/N and contamination by a sky line. As the ratio of [\ion{N}{II}] $\lambda$6548 to [\ion{N}{II}] $\lambda$6583 is constant, we do not lose any information by neglecting the [\ion{N}{II}] $\lambda$6548 line and in fact, the fitting quality improves when [\ion{N}{II}] $\lambda$6548 is dropped. We also note that the H$\alpha$ and [\ion{N}{II}] lines are always spectrally resolved. We probe a 12-dimensional parameter space defined by the broad- and narrow-component velocities, velocity dispersions, and line intensities of the four emission lines. Additionally, we experiment with fitting the four emission lines with a single-component model, i.e., a six-dimensional parameter space defined by four emission line intensities, one velocity, and one velocity dispersion. The single-component model corresponds to the physical scenario under which the lines share a common origin and exhibit symmetric profiles (e.g., from \ion{H}{II} regions or planar DIG). Similarly, for the blue data, we consider both two-component and single-component models for the observed H$\beta$ and [\ion{O}{III}] $\lambda$5007 lines. [\ion{O}{III}] $\lambda$4959 is not included in the fitting due to the low S/N.

To determine whether the spectra are better described by a two-component or single-component model, we use the Bayesian Information Criterion (BIC) \citep{BIC}. BIC measures the trade-off between model fit and complexity of the model. Models with lower BIC values are generally preferred. Note that here we compute BIC using only information from H$\alpha$ due to its high S/N. For each H$\alpha$ spectrum, we first compute BIC values for both the two-component fit ($\rm{BIC_{2comp}}$) and the single-component fit ($\rm{BIC_{1comp}}$). Then, we calculate $\Delta \rm{BIC} \equiv BIC_{\rm{2comp}} - BIC_{\rm{1comp}}$. After visually examining the spectra and the fitting results, we conclude that the two-component model is favored when $\Delta \rm{BIC}$ is $<-5$. We show in Figure~\ref{fig:example_spec} three example spectra along with their best-fit two-component models.

\section{Results} \label{sec:results}

The line-profile analysis described in \S\ 3 allows us to robustly decompose the observed emission signals into contributions from planar and extraplanar gas across the star-forming disks of NGC\,3511 and 3513.  In particular, the kinematics of the diffuse gas provide important information about how gas circulates at the disk-halo interface. The emission-line ratios in eDIG layers are known to be distinct from those of \ion{H}{II} regions \citep[e.g.,][]{Haffner2009}.
We discuss our findings for the two galaxies separately in the following subsections. 

\subsection{NGC\,3511}

Recall that NGC\,3511 is a disk galaxy of $\log\,\mstar/\msun=10.0$ with an inclination angle of $i = 74.5^{\circ}$ and ${\rm SFR}=0.8\,\msun\,{\rm yr}^{-1}$.  The IMACS observations have uncovered robust signals for multiple nebular lines across the disk of $\approx 27$ kpc in diameter. A distinct broad component is seen in approximately 15\% of the disk. For S1 data, the median value of the broad-component flux fraction is 0.19, with the 25th and 75th percentiles being 0.14 and 0.27, respectively. For S2 data, the flux fraction's percentile range spans from 0.26 to 0.42, with the median at 0.32. The kinematic and ionization properties of this broad component exemplify a typical eDIG layer, which are summarized in the following subsections. In addition, we also examine the connection of the eDIG layer to the local star formation properties along the disk.

\subsubsection{Gas Kinematics} \label{sec:ngc3511_kinematics}

We first examine the line-of-sight velocity dispersion ($\sigma$) and velocity centroid of the narrow, broad, and single-component models. The kinematic differences reveal distinctive physical properties between the narrow, broad, and single-component emission and constrain the origin of eDIG layers. Note that throughout the paper, we report $\sigma$ values after correcting for the instrumental line width ($\sigma^2=\sigma_{\rm obs}^2-\sigma_{\rm res}^2$).

\begin{figure}
    \centering
        \includegraphics[width=0.99\columnwidth]{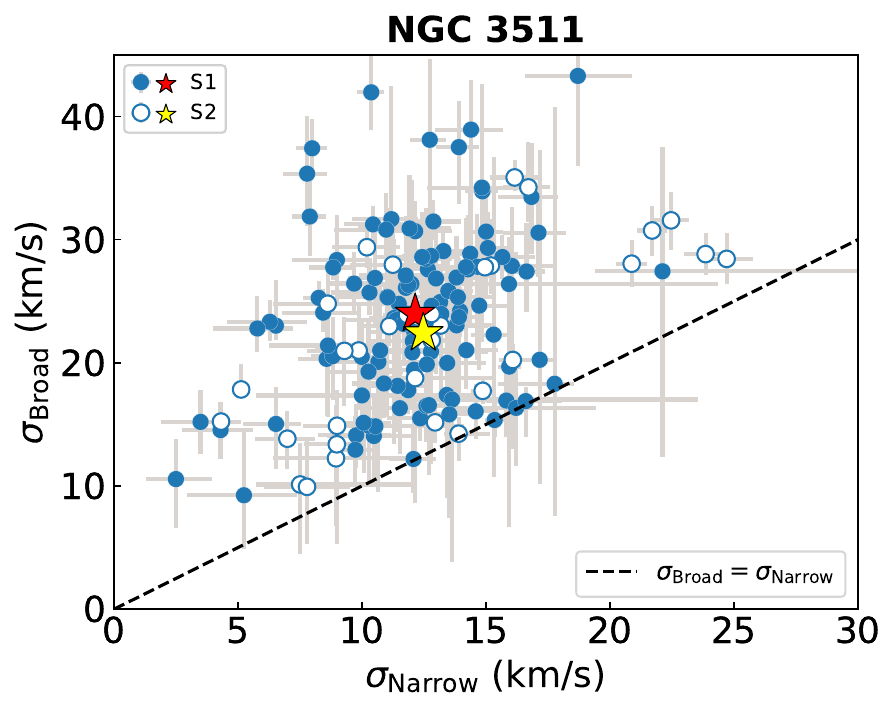} 
\caption{Comparisons of velocity dispersions between broad  ($\sigma_{\rm{Broad}}$) and narrow ($\sigma_{\rm{Narrow}}$) components, where $\sigma_{\rm{Broad}}$ are typically found to significantly exceed $\sigma_{\rm{Narrow}}$ with $\langle\sigma\rangle_{\rm{Broad}}\approx 24$ \kms\ and $\langle\sigma\rangle_{\rm{Narrow}}\approx 13$ \kms. The dashed line shows where $\sigma_{\rm{Broad}} = \sigma_{\rm{Narrow}}$. The median values from S1 and S2 are marked by the red and yellow star symbols, respectively. The much larger velocity dispersion measured in the broad component implies an origin in a thicker, extraplanar gaseous disk.}\label{fig:sigma_3511}
\end{figure}

\begin{figure*}
    \begin{minipage}{0.8\textwidth}
        \centering
        \includegraphics[width=\textwidth]{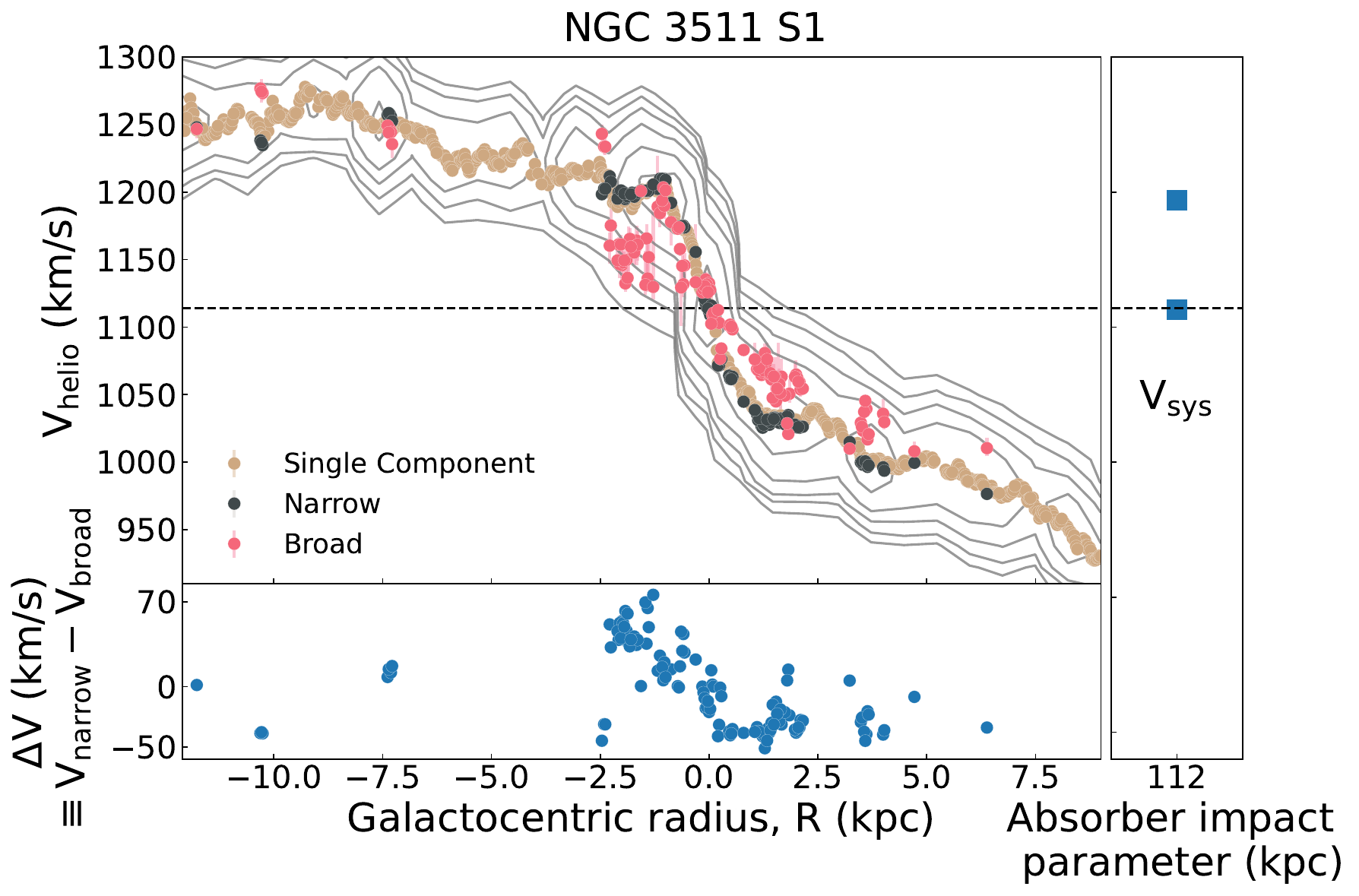}
    \end{minipage}\hfill
    \begin{minipage}{0.8\textwidth}
        \centering
        \includegraphics[width=\textwidth]{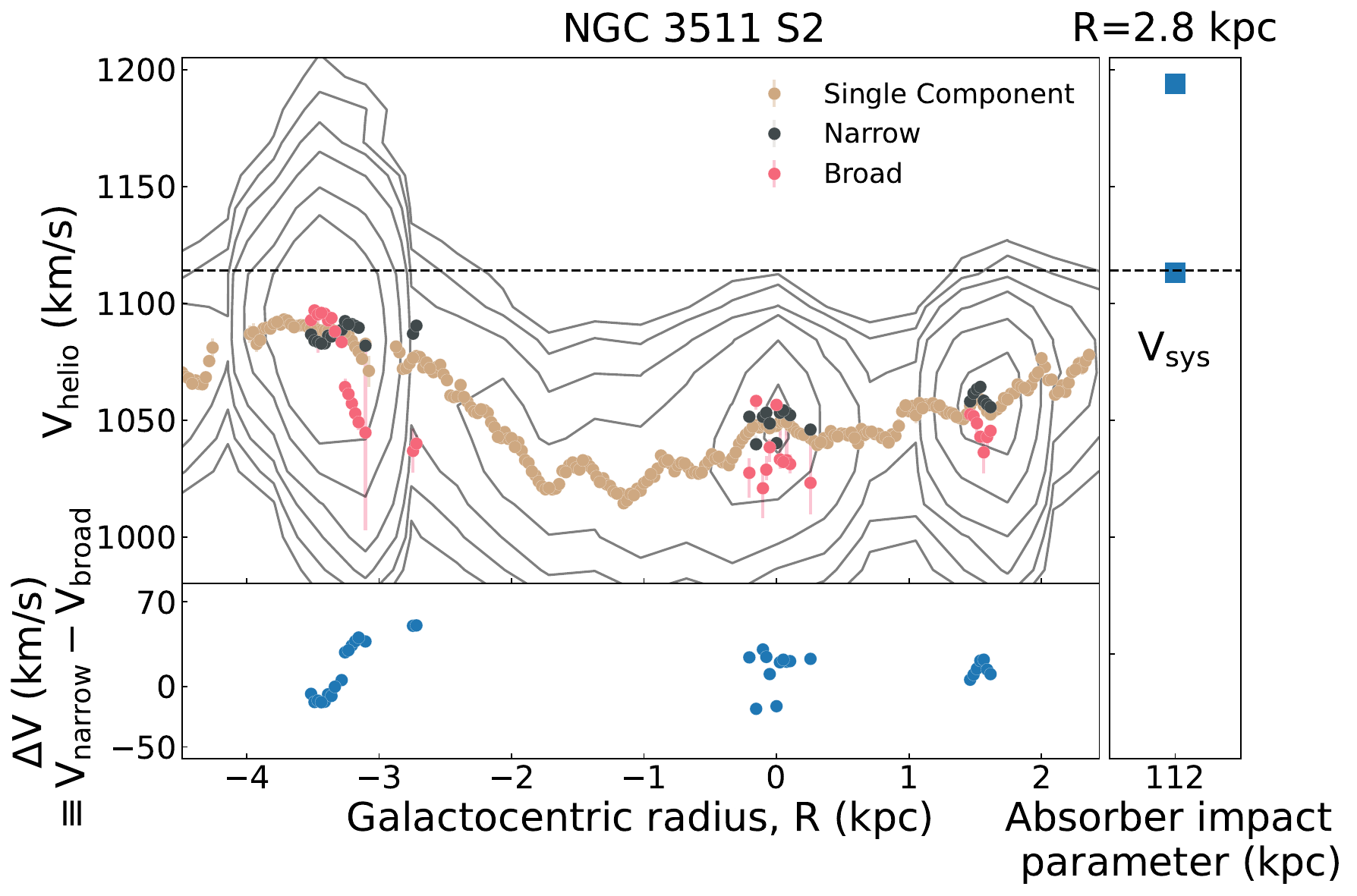}
\end{minipage}
\caption{Top: Radial velocity as a function of galactocentric radius for the narrow (black), broad (red), and single-components (yellow) along S1 in NGC\,3511. The contours indicate the constant H$\alpha$ flux levels at 2, 4, 8, 16, 40, and 80 times the average rms noise of the continuum in the final combined 2D spectral image along the position ($x$-) and velocity ($y$-) axes. We also show the difference in radial velocity between the narrow and broad components ($\Delta \rm V$). $|\Delta \rm V|$ ranges from 0 \kms to 70 \kms in projection, with a median value of $\langle|\Delta \rm V|\rangle = 34$ \kms. The broad component trends toward the systemic velocity compared to the narrow and single components. This rotational velocity lag is characteristic of eDIG. We also show the absorber velocities (squares), as compared to the systemic velocity of the galaxy (dashed line); most of the absorber column density ($\log\,N$(\ion{H}{I})/cm$^{-2} = 14.8$) is associated with the higher velocity component \citep{Keeney2017}. Bottom: Radial velocity as a function of the projected distance along S2 away from the disk plane ($z_{\text{proj}}$). We observe distinct, blueshifted velocities of the broad component compared to the disk, suggesting that they arise in local outflows.}\label{fig:velocity_curve_3511}
\end{figure*}

In Figure~\ref{fig:sigma_3511}, we compare the broad-component velocity dispersion with what is seen in the narrow component. Overall, we observe a much higher velocity dispersion for the broad emission than for the narrow or single-component emission. The median velocity dispersion for the broad emission is $\langle \sigma \rangle_{\text{Broad}} \approx 24$ \kms, whereas for the narrow emission we find $\langle \sigma \rangle_{\text{Narrow}} \approx 13$ \kms. Differences between S1 and S2 measurements are small. Furthermore, the broad emission shows a much larger scatter in the observed velocity dispersion, ranging from $\sigma_{\rm Broad} = $ 9 \kms to 45 \kms. (Note here that the thermal line width for $T \sim 10^{4}$ K gas is $\sim$ 10 \kms, so any measured value below this threshold is likely due to uncertainty in the observational data.) The larger velocity dispersion in the broad component implies an origin in a thicker gaseous disk, while the narrow- and single-component emission is likely from the thin galactic disk. In Section~\ref{sec:model}, we consider the scale height produced by the broad velocity dispersion and the implications for the extraplanar origin of the gas.

Next, we examine the line-of-sight velocity as a function of position along S1 and S2, which are displayed in the top and bottom panels of Figure~\ref{fig:velocity_curve_3511}, respectively.  The center of the galaxy is defined as the position where the rotation curve exhibits the steepest slope. We show S1 velocities as a function of galactocentric radius ($R$), and note that S2 crosses S1 at $R\,=\,2.8$ kpc. Since the galaxy is moderately inclined ($i\,=\,74.5^{\circ}$), we define the projected distance away from the disk plane in S2 ``$z_{\text{proj}}$'' to make a distinction from the intrinsic height ($z$) above the disk plane. In the top panel, we observe a clear difference in radial velocity between the narrow and broad components. This difference, $\Delta \rm V$, is shown as a function of $R$ below the rotation curve. 
The broad component is moving at a lower speed than the narrow or single component. The observed velocity lag ranges from $\Delta\,\rm V\,=\,0$ \kms to 70 \kms in projection, with a median value of $\langle\Delta\,\rm V\rangle\,=\,34$ \kms. A rotational velocity lag is characteristic of extraplanar gas, with typical values of $\approx 10 - 20$ km s$^{-1}$ kpc$^{-1}$ measured in both ionized and neutral phases in both edge-on and moderately inclined galaxies \citep[e.g.,][]{Heald06,Heald06b,Heald07,Bizyaev2017, Oosterloo2007,Gentile2013,Marasco2019}.

Due to the $n_{e}^{2}$ dependence of $I(H\alpha)$, we expect that the eDIG signal in NGC\,3511 is dominated by gas that is close to the disk ($|z| \lesssim 1$ kpc), and thus a lag as large as $\Delta \rm V = 70$ km s$^{-1}$ somewhat exceeds expectations (note that the inclination correction for rotational velocity is negligible for NGC\,3511). It is possible that blueshifted outflows contribute to the radial velocities, as hinted at by the somewhat larger values of $\Delta \rm V$ on the receding side of the galaxy than on the approaching side (see the top panel of Figure~\ref{fig:velocity_curve_3511}). We also observe evidence of outflows in S2, as shown in the bottom panel of Figure~\ref{fig:velocity_curve_3511}, where the velocity of the broad component is distinct and blue-shifted from the narrow component. 

\subsubsection{Emission-Line Ratios}
\label{sec:lineratios}

Emission-line ratios provide a diagnostic power for assessing the physical conditions of the gas.  We leverage different line ratios to determine the likelihood that the broad component indeed originates in an extraplanar gas layer.

\begin{figure*}
    \centering
        \includegraphics[width=\textwidth]{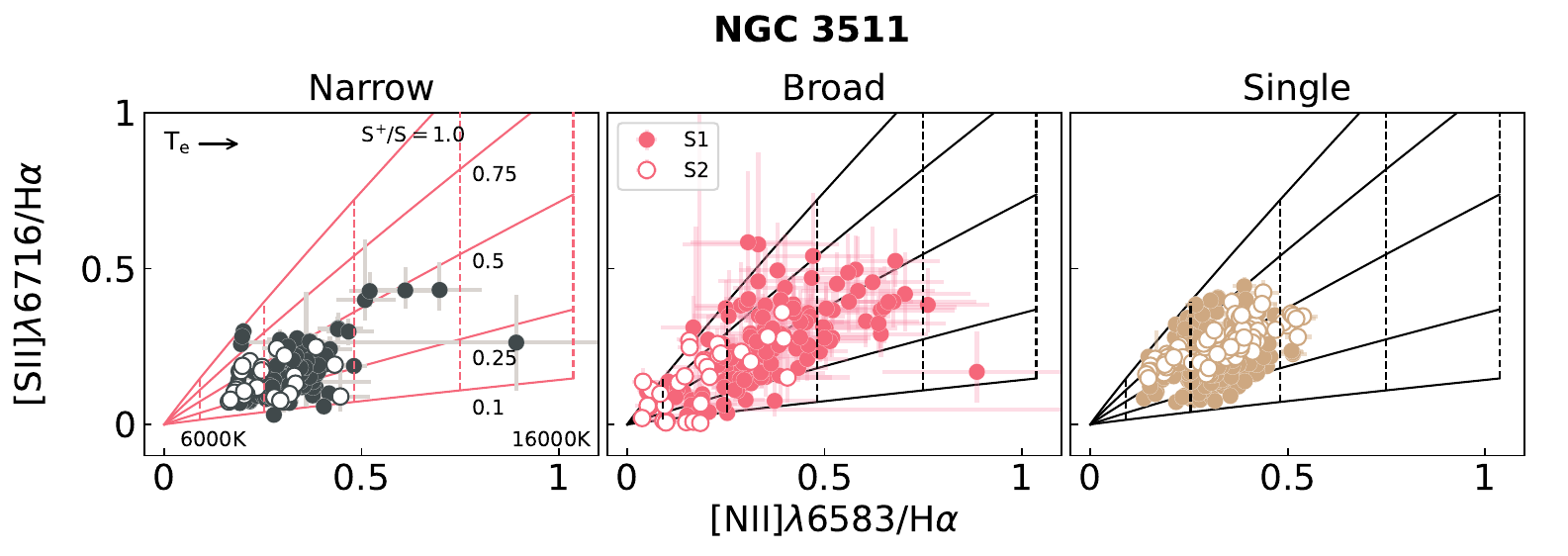} 
\caption{Nebular line ratios, [\ion{S}{II}]$\lambda$6716/H$\alpha$ versus [\ion{N}{II}]$\lambda$6583/H$\alpha$, for the narrow (black symbols in the {\it left} panel), broad (red in the {\it middle}), and single (yellow in the {\it right}) emission-line components in NGC\,3511 for regions with S/N$>10$ for all three emission lines ([\ion{S}{II}]$\lambda$6716, [\ion{N}{II}]$\lambda$6583 and H$\alpha$) to ensure the robustness of the reported line ratio values. Measurements along S1 are plotted with filled circles, while those along S2 are shown in open circles. The error bars show the 16th-84th percentiles of the measurement quantities. Under the assumption that NGC 3511 has a metallicity of $1/3$ solar, the solid lines correspond to a constant ionization fraction of ${\rm S}^{+}/{\rm S} = \{1.0,0.75,0.5,0.25,0.1\}$ from top to bottom, and the dashed lines indicate a constant electron temperature of $T_e = \{0.6,0.8,1.0,1.2,1.4,1.6\} \times 10^{4} \, \mathrm{K}$ from left to right. We see that the broad emission has a trend toward elevated line ratios compared to the narrow emission in both [\ion{S}{II}]$\lambda$6716/H$\alpha$ and [\ion{N}{II}]$\lambda$6583/H$\alpha$, corresponding to a higher temperature and a higher fraction of S$^+$ ions, which is consistent with observations of eDIG in other galaxies.}\label{fig:SII_NII_Ha_3511}
\end{figure*}

Physically, [\ion{N}{II}]$\lambda$6583/H$\alpha$ and [\ion{S}{II}]$\lambda$6716/H$\alpha$ probe the abundances, temperature and ionization state of the gas. The explicit dependence is shown in \citet{Haffner1999} and \citet{Osterbrock2006}: 
\begin{equation}
\frac{I([\ion{N}{II}]\lambda 6583)}{I({\rm H\alpha})} = 1.63 \times 10^{5}
\bigg(\frac{{\rm H}^{+}}{\rm H} \bigg)^{-1} \bigg(\frac{\rm N}{\rm H}\bigg)
\bigg(\frac{{\rm N}^{+}}{\rm N}\bigg) \times T_{4}^{0.426} \text{e}^{-2.18/T_{4}},
\label{eqn:NII_Ha}
\end{equation}
and
\begin{equation}
\frac{I([\ion{S}{II}]\lambda 6716)}{I({\rm H\alpha})} = 7.67 \times 10^{5}
\bigg(\frac{{\rm H}^{+}}{\rm H} \bigg)^{-1} \bigg(\frac{\rm S}{\rm H}\bigg)
\bigg(\frac{{\rm S}^{+}}{\rm S}\bigg) \times T_{4}^{0.307} \text{e}^{-2.14/T_{4}},
\label{eqn:SII_Ha}
\end{equation}
where $T_4 = \frac{T}{10^4 \, \rm{K}}$. We estimate the metallicity of NGC\,3511 to be $\sim 1/3$ solar by adopting a mass-metallicity relation appropriate for nearby galaxies \citep{Tremonti2004}.  For the solar abundance pattern, we expect N/H$ = 6.8 \times 10^{-5}$ and S/H$ = 1.3 \times 10^{-5}$ \citep{Asplund2009}, but we adopt a sulphur to nitrogen ratio that is twice the solar value following what is found in low-mass galaxies \citep[e.g.,][]{Rubin1984}. For the warm ionized gas, we assume that H and N are 100\% and 80\% ionized, respectively, based on measurements and models of the Milky Way ISM \citep{Reynolds1998, Sembach2000}. 
The ionization energy of N is 14.5 eV, while the ionization energy of S is 10.4 eV. This means that when N is mostly ionized, S is also mostly ionized. However, ${\rm S}^{+} \rightarrow {\rm S}^{2+}$ at 23.3 eV and  ${\rm N}^{+} \rightarrow {\rm N}^{2+}$ at 29.6 eV. Therefore, ${\rm S}^{+}/{\rm S}$ can vary while ${\rm N}^{+}/{\rm N}$ remains constant if the ionizing energy does not reach 29.6 eV. Consequently, we expect more variation in ${\rm S}^{+}/{\rm S}$ under eDIG conditions. 
 
In Figure~\ref{fig:SII_NII_Ha_3511}, we plot [\ion{S}{II}]$\lambda$6716/H$\alpha$ versus [\ion{N}{II}]$\lambda$6583/H$\alpha$ for the narrow, broad, and single-component emission detected in both S1 and S2.  We also include model expectations for constant electron temperature ($T_{e}$, dashed) and constant ${\rm S}^{+}/{\rm S}$ (solid) in each panel. 

Our observed single and broad-component line ratios are consistent with arising from \ion{H}{II} regions and the DIG, respectively. The single-component emission has intermediate line ratios, suggesting an origin with a mixture of physical conditions. Indeed, we see in Figure~\ref{fig:SII_NII_Ha_3511} that the broad component tends toward a lower ionization state (higher ${\rm S}^{+}/{\rm S}$) compared to the narrow component; this is observed in diffuse gas in the Milky Way and nearby galaxies and is due to the lower ionization parameter in the DIG and the reprocessing of the ionizing spectrum as it travels through the ISM (see \citealt{Haffner2009} and references therein). Additionally, we see in Figure~\ref{fig:SII_NII_Ha_3511} that the elevated line ratios in the broad component correspond to warmer temperatures ($T_{e} \approx 1 - 1.2 \times 10^{4}$ K) than the narrow component. This change in physical conditions in the DIG compared to \ion{H}{II} regions suggests spectral reprocessing of the ionizing radiation field and, potentially, supplemental sources of heating. Supplemental heating mechanisms that may raise the electron temperature of the diffuse gas include cosmic ray heating \citep{Wiener2013} and dissipation of turbulence \citep{Minter1997}. The elevated line ratios may also arise due to photoionization by a hard spectrum produced by hot, low-mass evolved stars \citep{Flores-Fajardo2011}, shocks, or turbulent mixing layers \citep[e.g.,][]{Rand1998}.

\begin{figure}
    \centering
        \includegraphics[width=0.99\columnwidth]{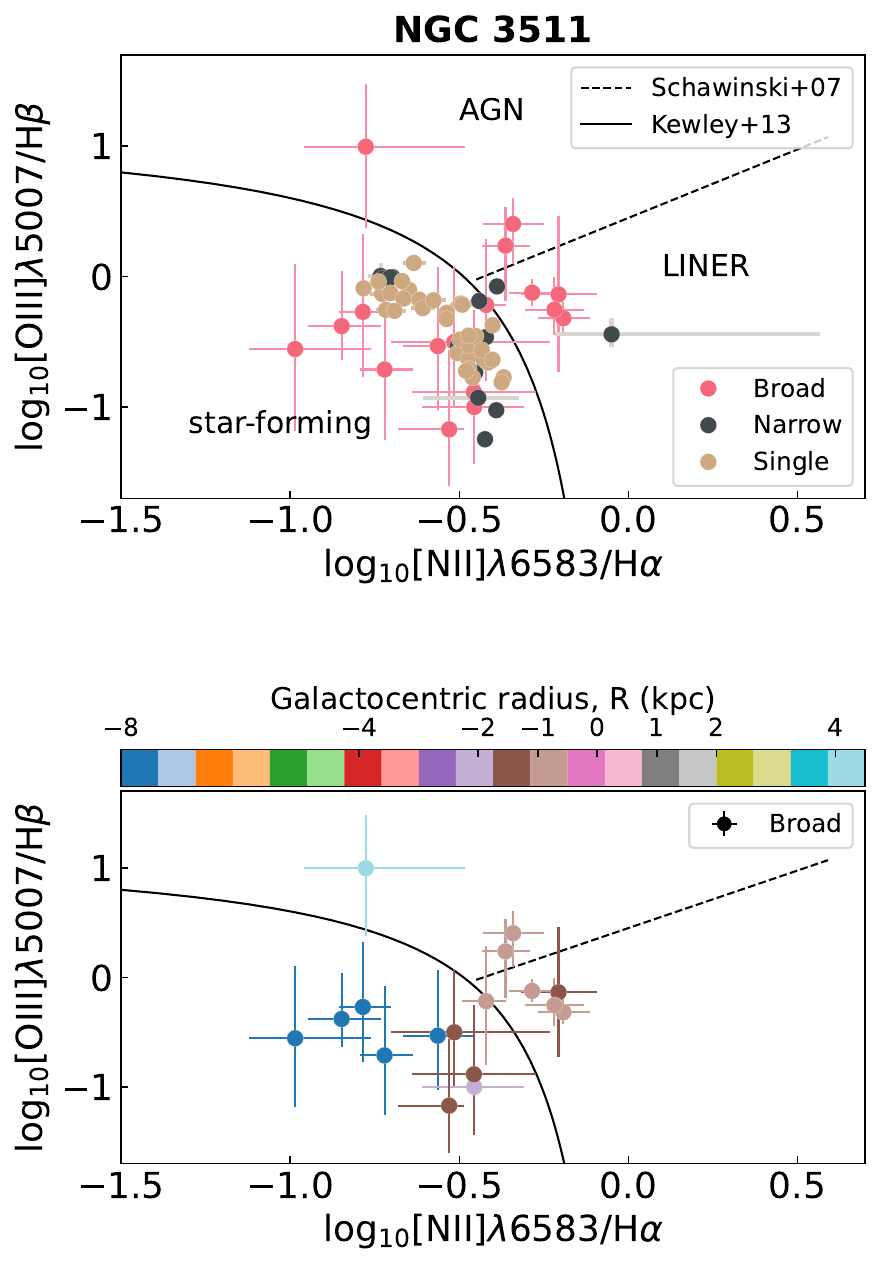} 
\caption{Top: BPT diagram showing line ratios of the narrow, broad, and single emission-line components. The color schemes for the different components are the same as in Figure~\ref{fig:SII_NII_Ha_3511} and are consistent throughout the paper. Some of the broad emission falls above and to the right of the solid line from \citet{Kewley2013}, where ionization is {\it not} due to star formation. Most of this emission occupies the LINER region below the dashed line from \citet{Schawinski07}. While a few measurements fall in the region indicating ionization by an AGN, they are consistent with LINER-like emission or ionization by star formation within the errors. Bottom: BPT diagram of the broad components. The data points are color-coded based on their galactocentric radii. The proximity of the LINER-like emission to the galactic center suggests that the emission may come from gas heated by shocks from a nuclear outflow.}\label{fig:BPT_3511}
\end{figure}

To probe the ionization mechanism producing the DIG layer, we now include the blue data and consider [\ion{O}{III}]$\lambda$5007/H$\beta$ versus [\ion{N}{II}]$\lambda$6583/H$\alpha$ in Figure~\ref{fig:BPT_3511}. Note that although there is an offset of one slit width between the blue and red data, the spatial fluctuations observed along S2, perpendicular to S1, indicate that strong line ratios vary on scales much larger than $0.7''$. Therefore, we assume the same gas kinematics for [\ion{O}{III}]$\lambda$5007 and H$\beta$ as for [\ion{N}{II}]$\lambda$6583 and H$\alpha$; we only fit the amplitudes of [\ion{O}{III}]$\lambda$5007 and H$\beta$. Figure~\ref{fig:BPT_3511} is commonly referred to as the BPT diagram and is used to distinguish between gas photoionized by AGN and by young stars in \ion{H}{II} regions \citep{Baldwin1981}. Gas with line ratios below the \citet{Kewley2013} line is expected to originate in star-forming regions. \citet{Schawinski07} provide a division line between Seyferts (upper) and low-ionization nuclear emission-line regions (LINERs; lower). We show that most of the narrow and single-component emission falls in the star-forming region; this is consistent with the conclusion that this emission arises in and around \ion{H}{II} regions. On the other hand, some of the broad components have higher line ratios in both [\ion{O}{III}]$\lambda$5007/H$\beta$ and [\ion{N}{II}]$\lambda$6583/H$\alpha$ than gas in the \ion{H}{II} regions. We color-code the broad components based on their distance from the galactic center in the bottom panel of Figure~\ref{fig:BPT_3511}. It is clearly shown that the broad components sitting to the right of the \citet{Kewley2013} line tend to reside near the galaxy center. The proximity of the LINER-like emission to the nucleus suggests that this emission may arise from gas heated by shocks from a nuclear outflow. While it is possible that a low-luminosity AGN contributes to photoionizing the nuclear region of this Seyfert galaxy, the LINER-like emission falls at $R \gtrsim 0.5$ kpc, leading us to favor a shock origin for the elevated line ratios.

\begin{figure*}
    \centering
        \includegraphics[width=0.825\textwidth]{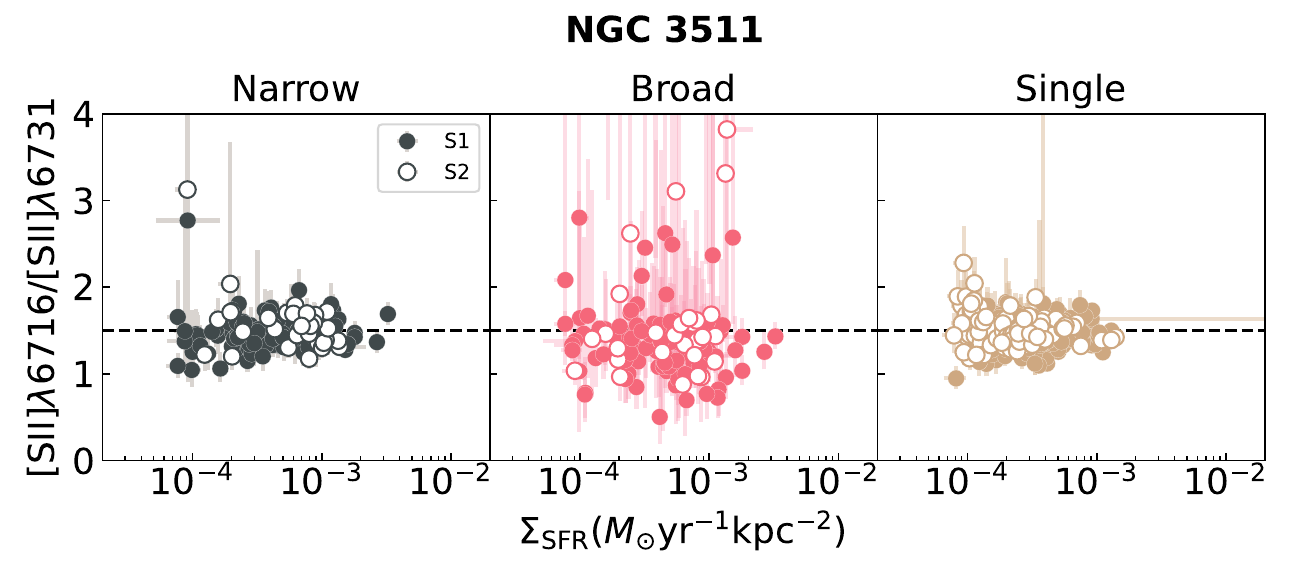} 
\caption{[\ion{S}{II}]$\lambda$6716/[\ion{S}{II}]$\lambda$6731 versus SFR surface density ($\Sigma_{\rm{SFR}}$) for the narrow (left), broad (middle), and single-component (right) emission spectra in NGC\,3511. The [\ion{S}{II}]$\lambda$6716/[\ion{S}{II}]$\lambda$6731 line ratio probes gas electron density. The narrow and single-component spectra have line ratios around the low-density limit (indicated by the dashed line), which corresponds to an electron density of $n_{e} \leq 10 \, \rm cm^{-3}$. The broad-component spectra display a larger scatter and have generally lower [\ion{S}{II}]$\lambda$6716/[\ion{S}{II}]$\lambda$6731 values with a median of $\approx$1.2, corresponding to an electron density of $n_{e} \approx 200 \, \rm cm^{-3}$. The higher average electron density in the broad component suggests that the gas originates from dense shells and filaments. Additionally, for all components, the line ratio is independent of $\Sigma_{\rm{SFR}}$, indicating that the gas density is independent of proximity to photoionizing sources.}\label{fig:SII_SFR_3511}
\end{figure*}

We also infer the underlying electron gas density based on the observed [\ion{S}{II}]$\lambda$6716/[\ion{S}{II}]$\lambda$6731 doublet ratios. In Figure~\ref{fig:SII_SFR_3511}, we compare [\ion{S}{II}]$\lambda$6716/[\ion{S}{II}]$\lambda$6731 of the narrow, broad, and single-component emission. This ratio decreases as the electron gas density increases; at the low-density limit, [\ion{S}{II}]$\lambda$6716/[\ion{S}{II}]$\lambda$6731 = 1.5, corresponding to an electron density of $n_{e} \lesssim 10 \, \rm cm^{-3}$ \citep{Osterbrock2006}. The low-density limit is plotted with a dashed line for comparison. It is clear that the narrow and single components have line ratios close to the low-density limit (corresponding to $n_{e} \sim$ 10 $\rm cm^{-3}$ or lower). \ion{H}{II} regions are observed to have a broad range of electron densities ($n_{e} \sim 10 - 10^{3}$ cm$^{-3}$), although the values inferred from [SII] doublet ratio measurements are often on the order of $n_{e} \sim 100$ cm$^{-3}$ \citep[e.g.,][]{Hunt2009}. Consequently, the observed $n_{e} \lesssim 10$ cm$^{-3}$ is towards the low end of the range typically seen in nearby galaxies. It is plausible that emission from a more diffuse planar DIG also contributes to the narrow- and single-component spectra, in addition to emission from \ion{H}{II} regions.

The broad components, on the other hand, display a larger scatter and generally have lower [\ion{S}{II}]$\lambda$6716/[\ion{S}{II}]$\lambda$6731 values, with a median value of $\approx$ 1.2, implying $n_{e} \approx 200\,\text{cm}^{-3}$. Similar [\ion{S}{II}]$\lambda$6716/[\ion{S}{II}]$\lambda$6731 line ratios of the eDIG components are found in regions of M83 \citep{Boettcher2017}. The low doublet ratios and their implied high gas density are consistent with some of the broad components originating in shocks. 

\begin{figure*}
    \centering
        \includegraphics[width=0.825\textwidth]{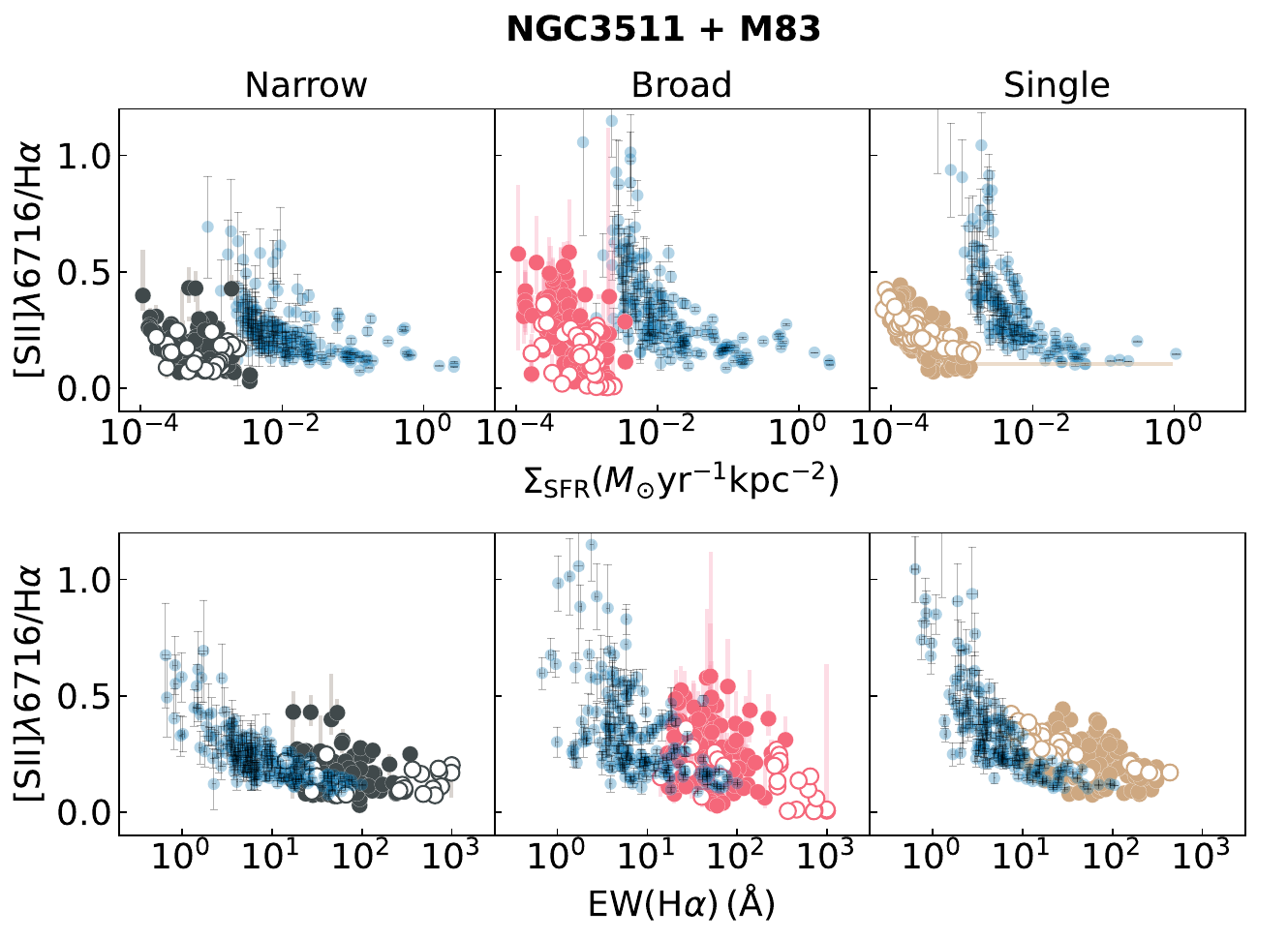} 
\caption{Top: [\ion{S}{II}]$\lambda$6716/H$\alpha$ line ratio versus $\Sigma_{\rm{SFR}}$ for the narrow ({\it left}), broad ({\it middle}), and single ({\it right}) components. Following Figure \ref{fig:SII_NII_Ha_3511}, measurements along S1 are plotted with filled circles, while those along S2 are shown in open circles. We overplot equivalent measurements for M83 from \citet{Boettcher2017} in blue. Bottom: The line ratios versus H$\alpha$ equivalent width (EW). The more coherent trend in the bottom panel suggests that the local EW is more predictive of the physical conditions in the gas than $\Sigma_{\rm{SFR}}$. This implies that the build-up of the eDIG layer and its properties are driven locally by the burstiness of the star formation history in the galaxy.}\label{fig:recombination_line_compare}
\end{figure*}

\subsubsection{Correlation with Star Formation Activities} \label{sec:3511_star_formation}

We have shown that both the kinematic properties (velocity lag and dispersion) and strong line ratios observed in the broad component support the presence of an eDIG layer in NGC\,3511.  To gain a better understanding of the physical processes that result in the eDIG layer, here we examine how the conditions in the eDIG layer are connected to local star formation. To quantify local star formation activities, we employ two empirical measures: the SFR surface density ($\Sigma_{\rm{SFR}}$) in units of SFR per unit area and the equivalent width of H$\alpha$ (EW).  The former provides an estimate of instantaneous local star formation intensity, while the latter provides an estimate of the burstiness of the star formation history.

We compute $\Sigma_{\rm{SFR}}$ from the observed H$\alpha$ flux, $F$(H$\alpha$) following
$\Sigma_{\text {SFR}}/(\mathrm{M}_{\odot} / \mathrm{yr} / \mathrm{kpc}^2)=7.9 \times 10^{-42} F(\mathrm{H} \alpha)/(\mathrm{ergs} / \mathrm{s} / \mathrm{kpc}^2)$ \citep{Kennicutt1998}. This quantity is directly proportional to the number of ionizing photons produced by stars per unit area. Since the lifetime of a typical O or B star is a few $\times 10^{6}$ years, $\Sigma_{\rm{SFR}}$ is a measure of the current rate of star formation over this timescale. A higher value of $\Sigma_{\rm{SFR}}$ suggests that the region has a higher density of, or is in closer proximity to, ionizing sources. Figure~\ref{fig:SII_SFR_3511} shows that the electron density of either the planar or extraplanar gas does not depend sensitively on its distance to the ionizing sources. 

In Figure~\ref{fig:recombination_line_compare}, we plot [\ion{S}{II}]$\lambda$6716/H$\alpha$ versus $\Sigma_{\rm{SFR}}$ in the top panel and EW in the bottom. We observe a clear decreasing trend with increasing $\Sigma_{\rm{SFR}}$ for the narrow and single components. This trend is characteristic of emission from \ion{H}{II} regions and planar DIG \citep[e.g.,][]{Haffner1999}, due to the lower ionization parameter in regions with lower SFRs and the reprocessing of the ionizing spectrum with distance from ionizing sources. For the broad, eDIG component, the scatter is large within the observed range in $\Sigma_{\rm{SFR}}$, and no correlation between the line ratio and $\Sigma_{\rm{SFR}}$ can be detected. However, when we examine the measurements from \citet{Boettcher2017} for M83 (shown in blue), which span a wider range of $\Sigma_{\rm{SFR}}$, we see a clearer decreasing trend with increasing $\Sigma_{\rm{SFR}}$ for all components. At fixed $\Sigma_{\rm{SFR}}$, we observe higher values of the emission-line ratios in M83 than in NGC\,3511. This result suggests that $\Sigma_{\rm{SFR}}$ is correlated with, but not fully predictive of, the physical conditions in the gas. 

In the bottom panel of Figure~\ref{fig:recombination_line_compare}, we show the line ratios as a function of the H$\alpha$ EW. The EW is an observational proxy for the SFR per unit mass with lower and higher EWs associated with on average older and younger stellar populations, respectively. In the bottom panel of Figure~\ref{fig:recombination_line_compare}, we observe an even clearer decreasing trend in line ratios with increasing EW, as compared to $\Sigma_{\rm{SFR}}$. This trend is coherent across the dynamic range in EW observed in M83 and NGC\,3511, suggesting that the EW is more predictive of the physical conditions in the gas than $\Sigma_{\rm{SFR}}$ alone. In NGC\,3511, the highest EWs (youngest stellar populations) are seen in S2 (open symbols in Figure \ref{fig:recombination_line_compare}).  The broad-component emission-line ratios in this region most closely resemble those in \ion{H}{II} regions. In contrast, NGC\,3511 S1 (closed symbols in Figure \ref{fig:recombination_line_compare}) and M83 generally have intermediate and low EWs, respectively, with increasingly elevated broad emission-line ratios observed with decreasing EW. Regions with the lowest EWs have broad emission-line ratios that are most characteristic of typical eDIG layers. This implies that the build-up of the eDIG layer, and the physical conditions in the gas, are driven locally by the burstiness of the galaxy's star formation history.

Additionally, we report that there is a lack of strong correlation between the broad-component flux fraction and $\Sigma_{\rm{SFR}}$ or EW. Specifically, for S1, the Spearman correlation coefficients are 0.03 with $\Sigma_{\rm{SFR}}$ and -0.12 with EW. For S2, these correlations are slightly stronger, at 0.16 with $\Sigma_{\rm{SFR}}$ and 0.27 with EW.

\begin{figure}
    \centering
        \includegraphics[width=0.99\columnwidth]{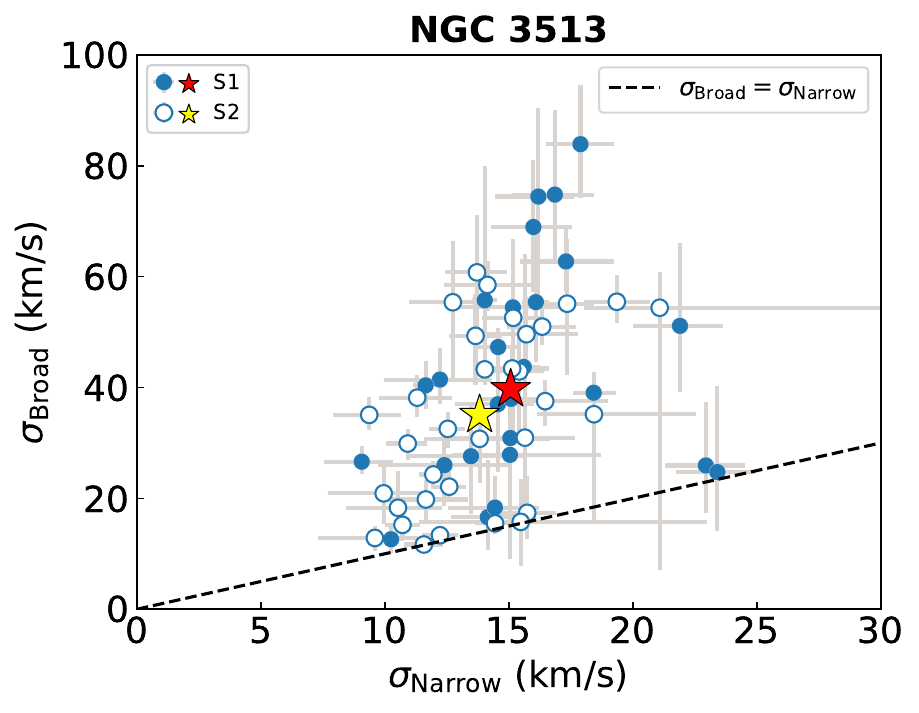} 
\caption{$\sigma_{\rm{Broad}}$ versus $\sigma_{\rm{Narrow}}$ for NGC\,3513. The symbols are the same as in Figure~\ref{fig:sigma_3511}, and the dashed line shows where $\sigma_{\rm{Broad}} = \sigma_{\rm{Narrow}}$. The median $\langle \sigma \rangle_{\rm{Broad}} = 40$ \kms, whereas $\langle \sigma \rangle_{\rm{Narrow}} = 15$ \kms. We note here that NGC\,3513 has higher observed broad-component velocity dispersions than NGC\,3511.}\label{fig:sigma_3513}
\end{figure}

\subsection{NGC\,3513}
\begin{figure*}
    \begin{minipage}{0.8\textwidth}
        \centering
        \includegraphics[width=\textwidth]{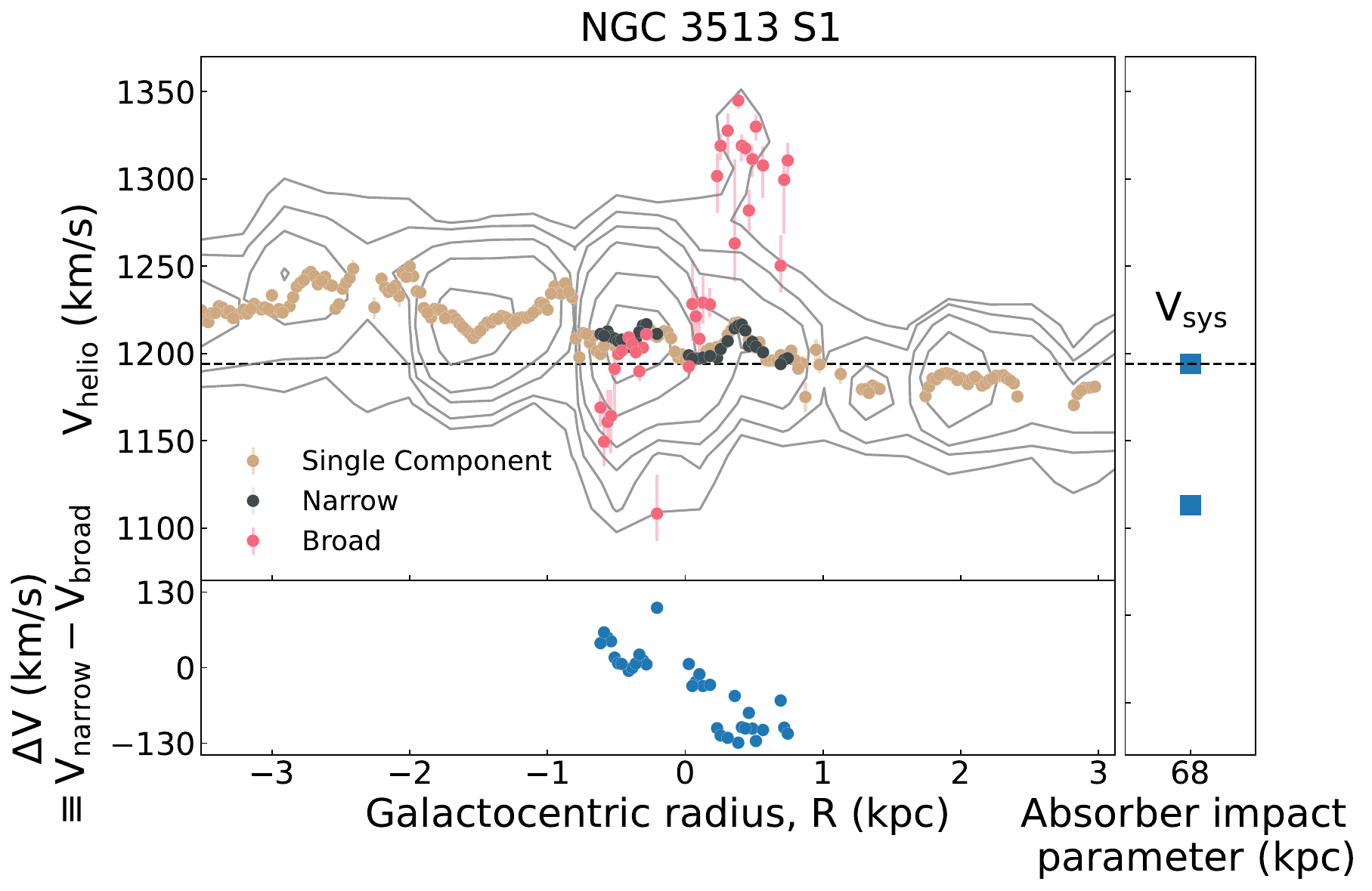}
    \end{minipage}\hfill
    \begin{minipage}{0.8\textwidth}
        \centering
        \includegraphics[width=\textwidth]{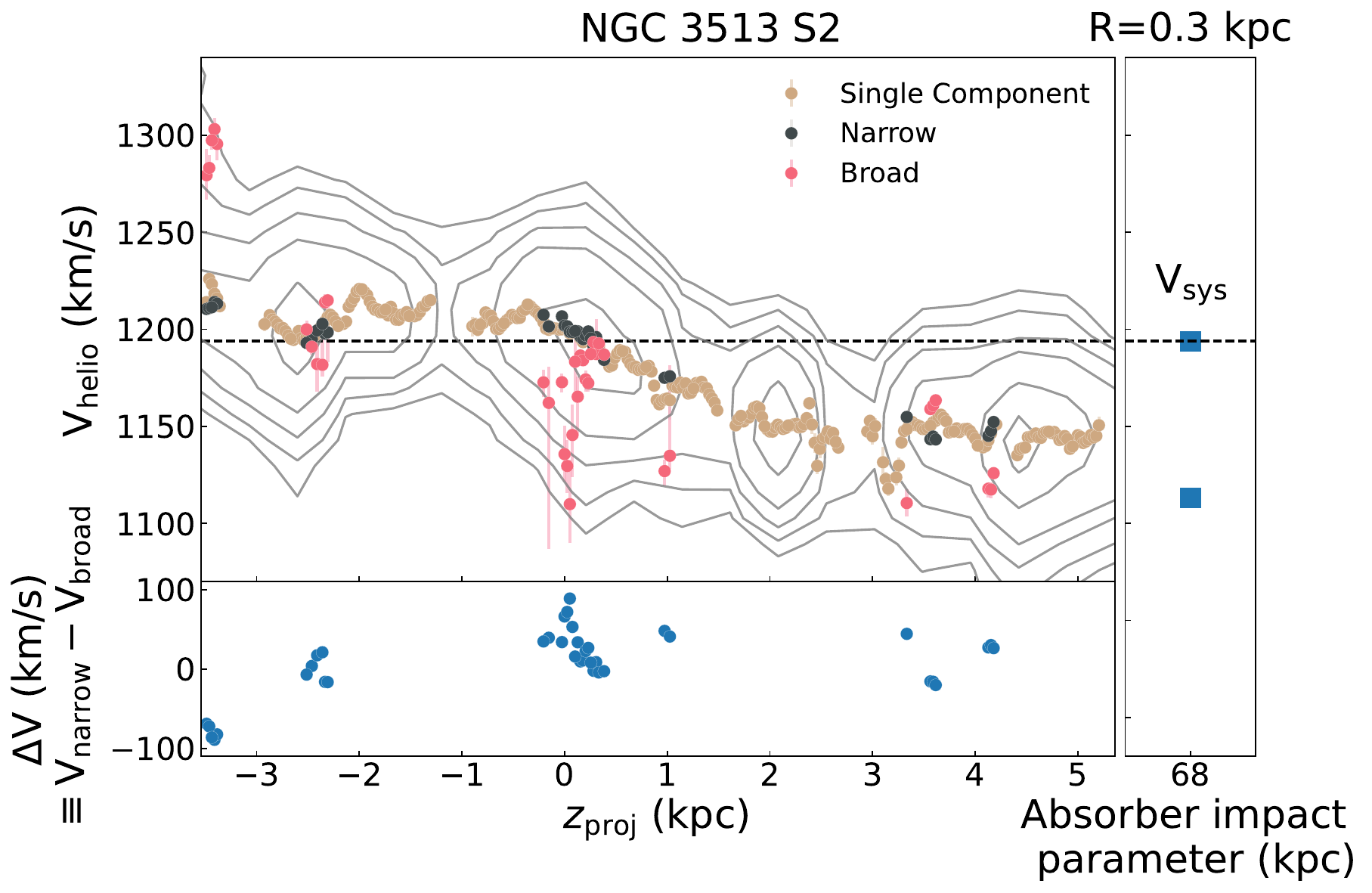}
\end{minipage}
\caption{The same as Fig.~\ref{fig:velocity_curve_3511} but for S1 (top) and S2 (bottom) observations of NGC\,3513. The contours show the H$\alpha$ flux levels in the final combined 2D spectral image, with position and velocity along the $x$ and $y$ axes, respectively. The contour levels are 2, 4, 8, 16, 40, and 80 times the average rms noise of the continuum. The broad component shows a significant velocity offset with respect to the disk, with $|\Delta \rm V| \leq 130$ km s$^{-1}$. In contrast to the widespread, lagging eDIG layer seen in NGC\,3511, the clumpy spatial distribution and kinematics of the broad component in NGC\,3513 indicate local outflows below the escape velocity.}\label{fig:velocity_curve_3513}
\end{figure*}

We now turn to NGC\,3513. Recall that NGC\,3513 is a companion galaxy of $\log\,\mstar/\msun=9.2$ with an inclination angle of $i = 39^{\circ}$ and ${\rm SFR}=0.2\,\msun\,{\rm yr}^{-1}$.  The IMACS observations have uncovered robust signals for multiple nebular lines across the disk of $\approx 8$ kpc in diameter. In Figure~\ref{fig:sigma_3513}, we compare the broad-component velocity dispersion with what is seen in the narrow component. Similar to NGC\,3511, we observe a much higher velocity dispersion for the broad emission than for the narrow or single-component emission. The median velocity dispersion for the broad emission is $\langle \sigma \rangle_{\text{Broad}} \approx 40$ \kms, whereas for the narrow emission we find $\langle \sigma \rangle_{\text{Narrow}} \approx 15$ \kms. Differences between S1 and S2 measurements are small.   The line-of-sight velocity as a function of position along S1 and S2 displayed in Figure~\ref{fig:velocity_curve_3513} shows that a distinct broad component is seen in approximately 20\% of the disk. The median broad-component flux fraction value is 0.28 for S1 data, with the 25th and 75th percentiles being 0.21 and 0.40. The median is 0.27 for the S2 data, and the 25th and 75th percentiles are 0.15 and 0.41.  We note that for NGC\,3513, S2 crosses S1 at $R\,=\,0.3$ kpc. 

For this galaxy, the most prominent feature is the large velocity difference ($\Delta \rm V$) between the broad components and the narrow and single components, as shown in Figure~\ref{fig:velocity_curve_3513}. The escape velocities for NGC\,3513 at $R \in [-3,3]$ kpc are $\sim$\,300 \kms, based on the galactic gravitational potential model that we describe in Appendix~\ref{sec:mass_model}. The largest $\Delta \rm V$ value we observe in the broad component is $\sim$\,130 \kms ($\sim$\,207 \kms after correcting for inclination), which is smaller than the escape velocities. However, it is clear that the rotation curve displays a signature of gas outflow close to the center of the galaxy, a sign of disk-halo circulation. In addition, we observe only modest evidence for elevated line ratios in a minority of the broad components in NGC\,3513. We include the line ratio plots for reference in Appendix~\ref{sec:3513_line_ratio}.

\section{A Dynamic Equilibrium Model of NGC\,3511}
\label{sec:model}
The unexpectedly large scale height of eDIG layers has puzzled researchers since its discovery. The characteristic scale height of the eDIG layer in the Milky Way and nearby edge-on disk galaxies is $h_{z} \sim$ 1 kpc, which is a factor of a few larger than the thermal scale height \citep[e.g.,][]{Rand1997, Haffner1999, Collins2001}. The scale height problem is central to our understanding of eDIG vertical structure and dynamics. In this section, we study the extent to which thermal and turbulent pressure gradients can support the eDIG layer in NGC\,3511 by examining the expected scale heights supported by different pressure forces, under the assumption of vertical hydrostatic equilibrium. 
Although it has been demonstrated that additional support from the magnetic field and cosmic ray pressure gradients may be needed in some galaxies \citep[e.g.,][]{Boettcher2016}, we do not have observational constraints to measure these non-thermal contributions in NGC\,3511. Instead, we ask whether the observed thermal and turbulent motions imply the need for additional non-thermal pressure support to produce the expected eDIG scale height. 

We note that even if vertical hydrostatic equilibrium does not hold in the eDIG, we expect the scale height of the layer to exceed the thermal value if $P_{\text{turb}}/P_{\text{th}}$ = $(\sigma_{\text{turb}}/\sigma_{\text{th}})^{2} \gtrsim 1$, where $P_{\text{turb}}$ and $P_{\text{th}}$ are the turbulent and thermal pressures in the gas, respectively, and $\sigma_{\text{turb}}$ and $\sigma_{\text{th}}$ are the corresponding velocity dispersions. Regardless of whether hydrostatic equilibrium is satisfied, the presence of significant, non-thermal pressure will cause the eDIG layer to expand to a larger scale height than expected from thermal pressure alone. In \S~\ref{sec:ngc3511_kinematics}, we find a median $\langle \sigma \rangle_{\text{turb}} = \sqrt{\langle \sigma \rangle_{\text{broad}}^{2} - \sigma_{\text{th}}^{2}} \approx 22$ \kms for the broad component in NGC\,3511, and therefore $P_{\text{turb}}/P_{\text{th}} \approx 5$. If the velocity dispersion is anisotropic and is largest perpendicular to the disk, as may arise in a galactic fountain, then the relative contributions of turbulent and thermal motions to the vertical pressure support may be even higher. This suggests that, even under non-equilibrium conditions, the vertical extent of the eDIG layer exceeds the thermal scale height in NGC\,3511. In the remainder of this section, we quantify the scale height that arises from thermal and turbulent pressure support in the case that hydrostatic equilibrium well describes the gas dynamics.

\begin{figure*}
    \centering
        \includegraphics[width=0.99\textwidth]{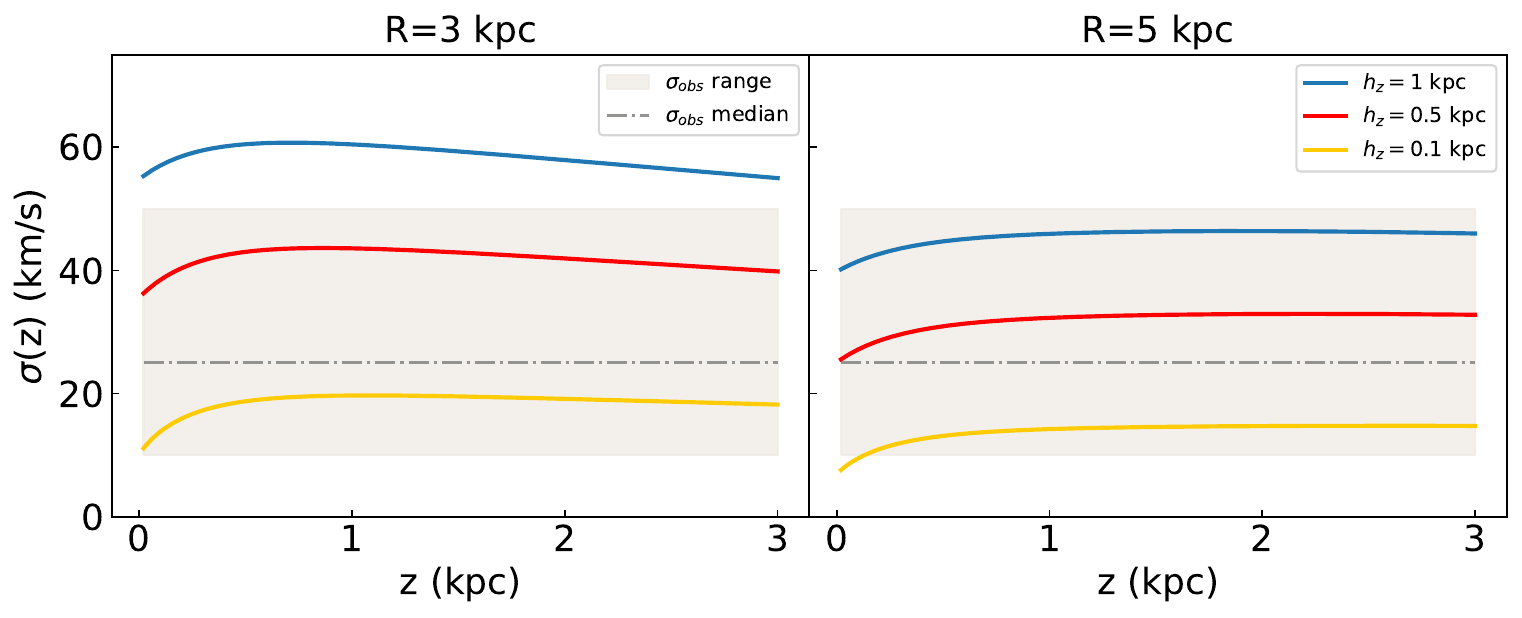} 
\caption{eDIG velocity dispersion ($\sigma$), predicted from our gravitational potential model for NGC 3511, as a function of height above the disk ($z$) for a characteristic set of exponential scale heights $h_{z} = \{0.1,0.5,1\}$ kpc, at the galactocentric radii $R=3$ kpc (left panel) and $R=5$ kpc (right panel). From $z \approx 0 - 3$ kpc, the variation of $\sigma$ with $z$ is small. Shaded regions show the range of observed $\sigma$ values in NGC 3511; the dash-dotted lines represent the median observed $\sigma$.}\label{fig:sigma_hz}
\end{figure*}

Under the hydrostatic equilibrium assumption, we expect the vertical pressure gradient to be balanced by gravity following 
\begin{equation} \label{eqn:equilibrium}
\frac{\partial P(z, R)}{\partial z} = - \rho(z, R) \frac{\partial \Phi(z, R)}{\partial z},
\end{equation}
where $\frac{\partial \Phi(z, R)}{\partial z}$ is the gravitational acceleration perpendicular to the plane, along the $z$-direction. By adopting an equation of state of the form
\begin{equation}
P(z, R)=\sigma^{2}(z) \rho(z, R),
\end{equation}
where $\sigma(z)$ is the velocity dispersion as a function of height and $\rho(z, R)$ is the density profile of the eDIG layer, we find a general solution to Equation~\ref{eqn:equilibrium}:
\begin{equation} \label{eqn:gensol}
\frac{\rho(z, R)}{\rho(0, R)}= \frac{\sigma^{2}(0)}{\sigma^{2}(z)} \text{exp} \bigg\{ - \int_{0}^{z} \bigg( \frac{\text{d}
    z'}{\sigma^{2}(z')} \frac{\partial \Phi(z', R)}{\partial z'} \bigg) \bigg\}.
\end{equation}

To examine the connection between the velocity dispersion and the spatial profile of the eDIG layer, we first consider an exponential form for the eDIG vertical density distribution, $\rho(z) = \rho(0)e^{-z/h_{z}}$, where $h_{z}$ is the scale height. This exponential distribution is motivated by the vertical H$\alpha$ intensity profiles observed in the eDIG layers of some nearby, edge-on galaxies \citep[e.g.,][]{Rand1990}. It follows that the left-hand side of Equation~\ref{eqn:gensol} reduces to $e^{-z/h_{z}}$.
By taking the derivative with respect to $z$ and rearranging terms on both sides, Equation~\ref{eqn:gensol} is further simplified to
\begin{equation}
\frac{d \sigma^{2}(z)}{d z}- \frac{1}{h_{z}} \sigma^{2}(z)=-\frac{\partial \Phi}{\partial z},
\end{equation}
with a solution of
\begin{equation} \label{eqn:finalsol}
\sigma^{2}(z)= e^{z/h_z}\int_{z}^{\infty} e^{-z^{\prime}/h_z} \frac{\partial \Phi}{\partial z^{\prime}} d z^{\prime}.
\end{equation}
Equation \ref{eqn:finalsol} illustrates how $\sigma(z)$ is correlated with $h_z$ of an exponential density profile for a given $\Phi$. To calculate the predicted $\sigma(z)$ for NGC\,3511, we compute its galactic gravitational potential based on a mass model constructed from combining known stellar and dark matter components.  See Appendix~\ref{sec:mass_model} for details.

In Figure~\ref{fig:sigma_hz}, we show the model-predicted eDIG velocity dispersion, $\sigma(z)$, for a characteristic set of $h_{z} = \{0.1,0.5,1\}$ kpc, at $R = 3$ kpc (left panel) and $R = 5$ kpc (right panel). The shaded region indicates the range of observed $\sigma$ values in NGC 3511, while the dash-dotted line represents the median observed $\sigma$. We see from Figure~\ref{fig:sigma_hz} that from the disk plane ($z\approx 0$ kpc) to $z = 3$ kpc, the variation of $\sigma$ with $z$ is small. In addition, when $\sigma(z)$ is comparable to thermal motions alone ($\approx 10$ \kms), the eDIG layer is vertically compact, with $h_{z} \lesssim 0.1$ kpc. In contrast, the observed range in $\sigma_{\rm broad}$ implies a scale height as large as $h_{z} \approx 0.7$ kpc at $R = 3$ kpc and $h_{z} \approx 1.1$ kpc at $R = 5$ kpc. The observed $\langle \sigma \rangle_{\text{Broad}}$ translates to $h_{z} \approx 0.2 - 0.4$ kpc over this range in $R$. The large scale height inferred for the eDIG together with the observed large velocity dispersion, exceeding thermal motions, highlights the important role that turbulence plays in supporting the eDIG layer.  However, a caveat is the presumed exponential density profile, which does not necessarily provide an accurate description of the eDIG profile close to the disk.

Next, we consider a second scenario in which the eDIG velocity dispersion is constant with $z$-height. This scenario has the benefit of enabling direct comparisons of expected and observed $\sigma$. In turn, it allows a density profile that departs from an exponential form. Returning to Equation~\ref{eqn:gensol}, a constant $\sigma(z)=\sigma$ leads to a density profile of
\begin{equation}\label{eqn:const_sig_rho}
\rho(z) = \rho(0)e^{-(\Phi(z) - \Phi(0))/\sigma^{2}},
\end{equation}
which characterizes the expected density profile of eDIG for a given $\sigma$ and $\Phi$.

\begin{figure*}
    \centering
        \includegraphics[width=0.99\textwidth]{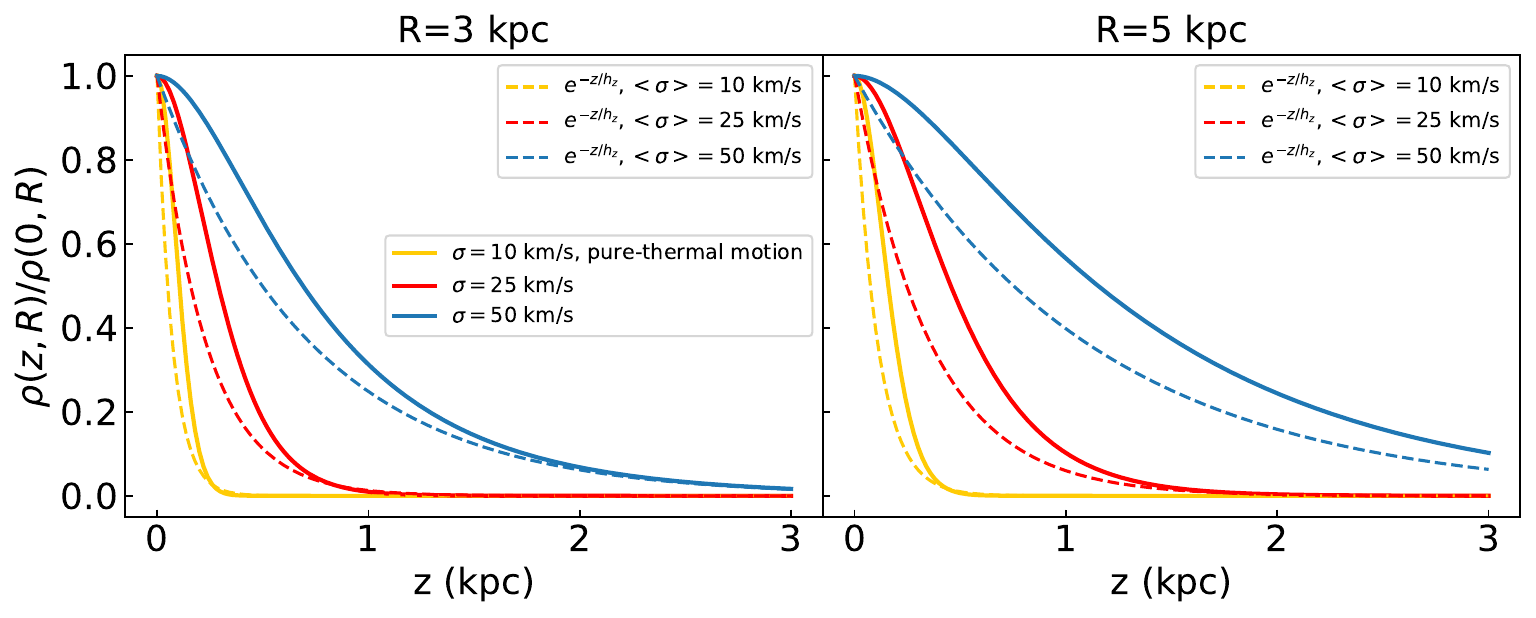} 
\caption{Normalized eDIG density profiles for two dynamic equilibrium models at the galactocentric radii $R=3$ kpc (left panel) and $R=5$ kpc (right panel), using a gravitational potential model representative of NGC\,3511. In the first model, the velocity dispersion, $\sigma(z)$, varies with $z$ according to Equation~\ref{eqn:finalsol} to produce an exponential density distribution (dashed lines). In the second model, $\sigma$ is constant with $z$ and the density distribution follows Equation~\ref{eqn:const_sig_rho} (solid lines). The profiles are shown for the observed range in $\sigma_{\text{Broad}}$ for NGC\,3511; the lower bound, median $\langle \sigma \rangle_{\text{Broad}}$, and upper bound are shown in yellow, red, and blue, respectively. For the exponential models, we show the profiles produced when the $n_{e}^2$-weighted average $\sigma(z)$ is equal to the indicated value, allowing direct comparison between the two models for a given observed $\sigma$. At small $z$, the eDIG layer is somewhat thicker when $\sigma(z) = \sigma$. Both models demonstrate that the observed non-thermal motions increase the thickness of the layer by at least a factor of a few compared to thermal motions alone (solid yellow curves).}\label{fig:density_profile}
\end{figure*}

In Figure~\ref{fig:density_profile}, we compare the exponential density profiles with those from Equation~\ref{eqn:const_sig_rho} at $R = 3$ kpc (left panel) and $R = 5$ kpc (right panel). The choices of constant $\sigma$ represent the lower bound, median, and upper bound of our observed $\sigma_{\text{Broad}}$ for NGC\,3511. For the exponential models, we show the density profiles produced with the $n_e^2$-weighted mean $\sigma(z)$ equal to the same set of constant $\sigma$ values. This $n_e^2$-weighting accounts for the density dependence of the emission measure from the observed H$\alpha$ intensity. We find that the inferred density distribution declines more gradually with $z$-height for the constant $\sigma$ models. For instance, the inferred eDIG density for the observed $\langle \sigma \rangle_{\text{Broad}}\approx 25$ \kms\ is $\approx 50$\% higher at $z = 0.5$ kpc from the constant $\sigma$ model than that from the exponential profile. Figure~\ref{fig:density_profile} shows that for a given observed velocity dispersion, the eDIG layer is expected to be more spatially extended if $\sigma(z)$ varies less with $z$-height. 

We emphasize here that, due to the inclination angle of NGC\,3511, the observed $\sigma$ value and the \textit{vertical} $\sigma$ in our theoretical models may not be the same quantity. In the limiting case that the velocity dispersion is isotropic, the observed $\sigma$ value is the true vertical velocity dispersion. In the contrasting limiting case, where the velocity dispersion is fully in the $z$-direction, then the observed $\sigma$ is smaller than the true vertical $\sigma$ by a factor of $\cos(i)$, where $i$ is the inclination angle of the galaxy ($\sigma_{\rm true} = \sigma_{\rm observed} / \cos(i) $). The characteristic $\sigma$ may thus fall between the median observed $\langle \sigma \rangle_{\text{Broad}} \approx 25$ \kms and the inclination-corrected value of $\approx$ 80 \kms. In the case of a highly anisotropic dispersion, the vertical extent of the eDIG layer may therefore significantly exceed that predicted by the observed $\langle \sigma \rangle_{\text{Broad}}$. 

\section{Discussion and Conclusions}
\label{sec:discussion}
We have demonstrated a technique to kinematically separate extraplanar warm ionized gas from planar signals in high-resolution optical spectra of the sub-$L_{*}$ disk galaxy pair NGC\,3511/3513. We found a widespread eDIG layer in NGC\,3511 and confirmed its extraplanar nature through its 1) lagging rotational velocity ($\langle \Delta \rm V \rangle = 34$ \kms), 2) higher velocity dispersion ($\langle \sigma \rangle_{\rm Broad} = 24$ \kms, as compared to $\langle \sigma \rangle_{\rm Narrow} = 13$ \kms), and 3) elevated [\ion{N}{II}]$\lambda$6583/H$\alpha$ and [\ion{S}{II}]$\lambda$6716/H$\alpha$ emission-line ratios with respect to the \ion{H}{II} regions. In NGC\,3513, we observe evidence of local, ionized outflows close to the nucleus ($|\Delta \rm V| \lesssim 130$ \kms). These are among the first galaxies with $i \lesssim 75^{\circ}$ in which the eDIG kinematics have been spatially resolved, permitting direct constraints on dynamical models of the gaseous, disk-halo interface.

The observed eDIG layers in NGC\,3511 and NGC\,3513 exhibit different spatial distributions. In NGC\,3511, the eDIG is more widespread than in NGC\,3513, with detected signals on the scale of several kiloparsecs from the galaxy center ($|R| \leq 5$ kpc). This is suggestive of the extensive, "classical" eDIG layers observed in galaxies like the Milky Way and the edge-on galaxy, NGC 891 \citep[e.g.,][]{Rand1996, Haffner2003}. In contrast, NGC\,3513 displays localized outflows that are largely confined to $|R| < 1$ kpc. The difference in the spatial distribution is consistent with the known diversity in extraplanar H$\alpha$ emission in nearby, edge-on galaxies, in which local patches or filaments of H$\alpha$ emission are more commonly observed than the pervasive eDIG layer observed in the Galaxy \citep[e.g.,][]{Rossa2003b}. Thus, the extraplanar gas properties in NGC\,3511 may more closely resemble those of the Milky Way, while those of NGC\,3513 may be more representative of nearby, low-mass galaxies.

As this is among the first efforts to measure a projection of the vertical velocity dispersion of the eDIG \citep[see also][]{Haffner2003, Fraternali2004, Boettcher2017, Li2021}, a main goal of this paper is to study how the eDIG layer is vertically supported. The characteristic line widths observed in the eDIG of both NGC\,3511 and NGC\,3513 are a factor of a few larger than the thermal line widths, demonstrating that turbulent motions are dynamically important. Based on the median velocity dispersion of the broad components, the turbulent pressure significantly exceeds the thermal pressure in the eDIG of both galaxies, with $P_{\text{turb}}/P_{\text{th}} \approx 5$ in NGC\,3511 and $P_{\text{turb}}/P_{\text{th}} \approx 15$ in NGC\,3513. We thus expect that in both galaxies, the eDIG layer will expand to a scale height that is larger than the thermal scale height, regardless of whether the gas is in dynamical equilibrium. We quantified the scale height under the assumption of vertical hydrostatic equilibrium in NGC\,3511, and we found that the median observed velocity dispersion suffices to produce significant values ($h_{z} \approx 0.2 - 0.4$ kpc at $R = 3 - 5$ kpc; see Section~\ref{sec:model}). This is somewhat smaller than the characteristic $h_{z} \approx 1$ kpc observed in extensive eDIG layers in galaxies like NGC 891 \citep[e.g.,][]{Rand1990}. However, due to the high inclination of NGC\,3511, the observed $\langle \sigma \rangle_{\text{Broad}}$ may be significantly smaller than the true vertical $\sigma$ if the velocity dispersion is anisotropic; thus, the scale height may be under-predicted. There is also evidence that magnetic field and cosmic ray pressure gradients may contribute significantly to elevating the scale height of eDIG \citep{Boettcher2016, Boettcher2019}. Our techniques can be applied to other galaxies and we expect more observational measurements to enable a more robust comparison to predictions from high-resolution numerical simulations \citep[e.g.,][]{Chan2022}.

Whether or not the velocity dispersion of the eDIG depends on the star-formation properties of the host galaxy is an important open question. With the present sample, we cannot robustly assess the connection between the velocity dispersion and the global specific SFR (sSFR; i.e., whether or not higher turbulent energy injection per unit mass in higher sSFR galaxies tends to produce larger eDIG velocity dispersions). We consider here five galaxies with $i \lesssim 75^{\circ}$ for which measurements of the eDIG velocity dispersion are available - the NGC\,3511/3513 pair, M83 \citep{Boettcher2017}, and NGC\,3982 and NGC\,4152 \citep{Li2021}. Among these galaxies, M83 has the highest SFR (SFR $= 4.2 \, M_\odot$ yr$^{-1}$) and stellar mass ($M_{\text{star}} = 10^{10.5} \, M_\odot$; \citealt{Leroy2021}), but its sSFR of log(sSFR/yr$^{-1}$) $= -9.9$ is comparable to that of NGC\,3513 and $\approx 60$\% higher than that of NGC\,3511. The median eDIG velocity dispersion in M83, $\langle \sigma \rangle_{\rm Broad} = 96$ \kms, is a factor of $2 - 3$ higher than that in NGC\,3513 ($\langle \sigma \rangle_{\rm Broad} = 40$ \kms) and a factor of $4$ higher than that in NGC\,3511 ($\langle \sigma \rangle_{\rm Broad} = 24$ \kms). NGC\,3982 and NGC\,4152 have similar sSFRs to NGC\,3513 (log(sSFR/yr$^{-1}$) $= -9.8$ and $= -9.5$, respectively; \citealt{Leroy2019}), and their eDIG velocity dispersions are $\sigma = 50 - 60$ \kms \citep{Li2021}. However, the determination of any trends in the velocity dispersion with the sSFR is limited by two important caveats. First, due to the relatively high inclination of NGC\,3511, the vertical velocity dispersion in this galaxy may be significantly higher than observed. Secondly, the dynamic range in sSFR is small (by a factor of four) among these five galaxies. Our study thus motivates measurements of the eDIG velocity dispersion in a statistically significant sample covering an increased dynamic range in the sSFR.  

In contrast to the global star-formation properties, we have strong statistics to probe correlations of the spatially resolved eDIG properties with the local star-formation activity within each galaxy. The strong anti-correlation between the eDIG emission-line ratios and the local H$\alpha$ EW shown in Figure~\ref{fig:recombination_line_compare} suggests that the conditions in the extraplanar gas are determined by the local star-formation history, with significant variation observed within individual galaxies. Indeed, Figure~\ref{fig:recombination_line_compare} demonstrates that the physical conditions that most resemble those in the classical eDIG (in particular, elevated $T_{e}$), are associated with regions with the lowest EWs (i.e., the oldest stellar populations). This suggests that the conditions in the eDIG are more reflective of the burstiness of the star-formation history than of the instantaneous star-formation rate. 

Notably, we do not observe a statistically significant correlation between the eDIG velocity dispersion and the local $\Sigma_{\rm{SFR}}$ or H$\alpha$ EW. This is qualitatively consistent with the expectations of galactic fountain models. If the eDIG is produced by a galactic fountain flow, then we expect the gas temperature and ionization state to couple more closely to the local, underlying stellar population than the gas kinematics.  This is because under the fountain models, extraplanar gas clouds can travel distances ranging from hundreds of parsecs to a thousand parsecs away from their originating star-forming regions \citep[e.g.,][]{Collins2002}, resulting in the decoupling of the eDIG kinematics and the local, underlying disk properties. At the same time, the eDIG temperature and ionization state would more closely reflect the local stellar population and the associated photoionizing spectrum, as the recombination time is significantly shorter than the cloud orbital time. Going forward, we hope to gain more insight into how eDIG properties and dynamics are related to star-formation activities by obtaining observational data for more galaxies that are less inclined than NGC\,3511 that (1) meet the SFR per unit area threshold of \citet{Rossa2003} and (2) increase the dynamic range in SFR and sSFR both locally and globally.

The warm ionized gas kinematics at the disk-halo interface have implications for the co-evolution of galaxies and their circumgalactic medium. In the NGC\,3511/3513 system, we find that gas circulation at the disk-halo interface is present in both galaxies. While the presence of the sightline along the minor axis of NGC\,3511 has led to the association of the absorber towards QSO PMN\,J1103$-$2329 with this galaxy \citep{Stocke2013}, the closer proximity of NGC\,3513 to the sightline, at a projected distance of $d=68$ kpc north of the QSO, as well as the evidence for active outflow from star-forming regions in the galaxy, make NGC\,3513 a more likely host for the absorber. We note that for the gas to travel a distance of 68 kpc the launch velocity at $R=5$ kpc needs to be $\sim$250 \kms ($\sim$160 \kms measured at the observed inclination), and our measured gas velocity is below this threshold ($\lesssim 130$ \kms). However, while we do not observe direct evidence for warm outflows that will reach the spatial scale of the absorber, the absorber may be associated with past epochs of more vigorous star-formation activity.  Alternatively, the gas may be the cooled product of a hot outflow launched at higher velocities. To conclude, this system illustrates the need to proceed with caution when associating absorbers to host galaxies in pairs or group environments. NGC\,3511 and NGC\,3513 are members of the MHONGOOSE deep \ion{H}{1} survey being conducted with MeerKAT \citep{Sardone2021}, which will map the neutral gas morphology and kinematics in these galaxies to unprecedented sensitivity. Paired with our nebular spectroscopy, this will provide a multi-phase window on the role of stellar feedback in mediating the disk-halo connection via turbulent gas motions.

\section*{Acknowledgements}

We thank the anonymous referee for useful comments that improved the clarity of the manuscript. We thank Jay Gallagher and Ellen Zweibel for helpful discussions that motivated this study, and Christy Tremonti for useful discussions about the role of metallicity. HZ acknowledges discussions with Nick Gnedin and Andrey Kravtsov. EB acknowledges partial support by NASA under award number 80GSFC21M0002.

\section*{Data Availability}

The data underlying this article will be shared on reasonable request to the authors.

\bibliographystyle{mnras}
\bibliography{main} 

\appendix
\section{NGC\,3513: Emission-line Ratios}
\label{sec:3513_line_ratio}

\begin{figure*}
    \centering
        \includegraphics[width=\textwidth]{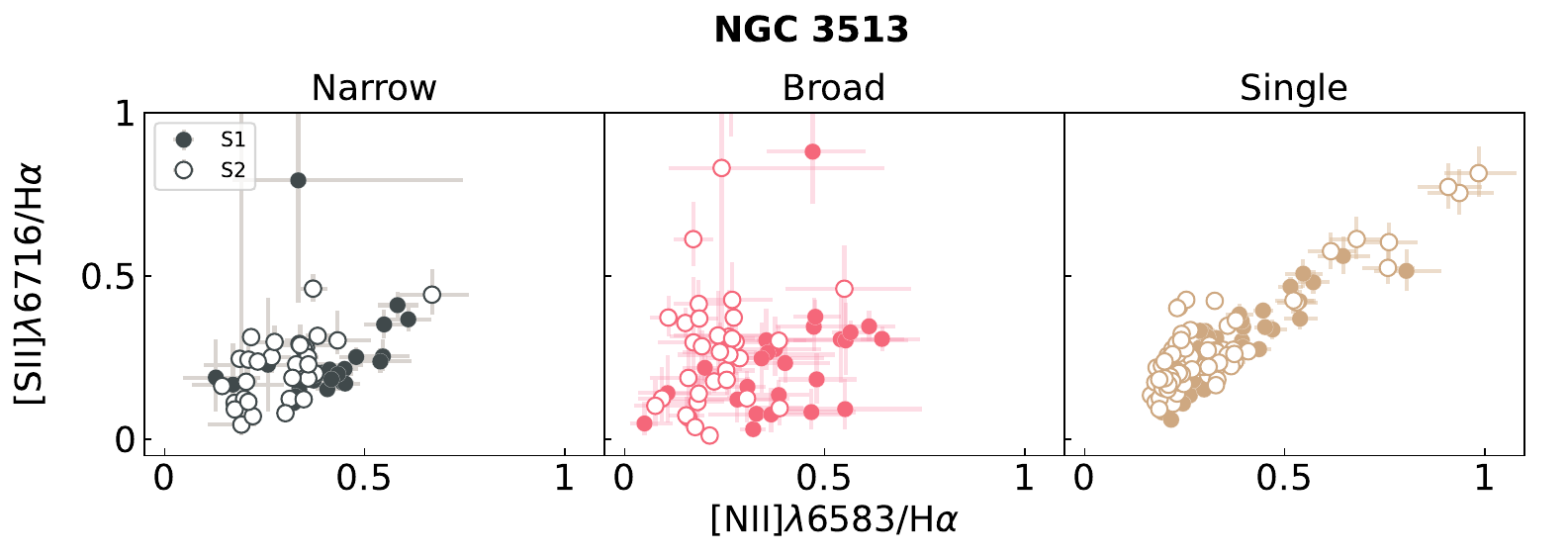} 
\caption{The same as Figure~\ref{fig:SII_NII_Ha_3511} but for NGC\,3513. The broad component does not occupy a distinct region of line-ratio parameter space with respect to the narrow component within the uncertainties.}\label{fig:SII_NII_Ha_3513}
\end{figure*}

Similar to Figure~\ref{fig:SII_NII_Ha_3511}, we show [\ion{S}{II}]$\lambda$6716/H$\alpha$ versus [\ion{N}{II}]$\lambda$6583/H$\alpha$ for NGC\,3513 in Figure~\ref{fig:SII_NII_Ha_3513}. The narrow-component emission-line ratios are similar to those in NGC\,3511, suggesting similar temperatures and ionization states in the \ion{H}{II} regions of both galaxies.  Additionally, while many of the single-component spectra have similar line ratios in both galaxies, we note that NGC\,3513 has a tail to elevated [\ion{S}{II}]$\lambda$6716/H$\alpha$ and [\ion{N}{II}]$\lambda$6583/H$\alpha \gtrsim 0.5$ that is absent in NGC\,3511. The elevated emission-line ratios imply a higher gas temperature in these regions, which are found near the outflows in the inner galaxy. As the velocity centroids of the single-component spectra largely trace the rotation of the disk (see Fig.~\ref{fig:velocity_curve_3513}), this gas does not appear to be part of an outflow itself. Instead, it may trace shocked, planar gas surrounding the star-forming regions that drive the outflows.

In contrast to NGC\,3511, NGC\,3513 does not show clear evidence for elevated emission-line ratios in the broad emission as compared to the narrow emission. The lack of a clear trend may be partially due to poorer statistics in NGC\,3513, where fewer spectra require a two-component model in this lower-mass galaxy. As the similar broad and narrow emission-line ratios suggest similar physical conditions in the gas, it is possible that the broad component primarily traces the base of the outflows where the temperature and ionization conditions are similar to those in the \ion{H}{II} regions.

\begin{figure*}
    \centering
        \includegraphics[width=\textwidth]{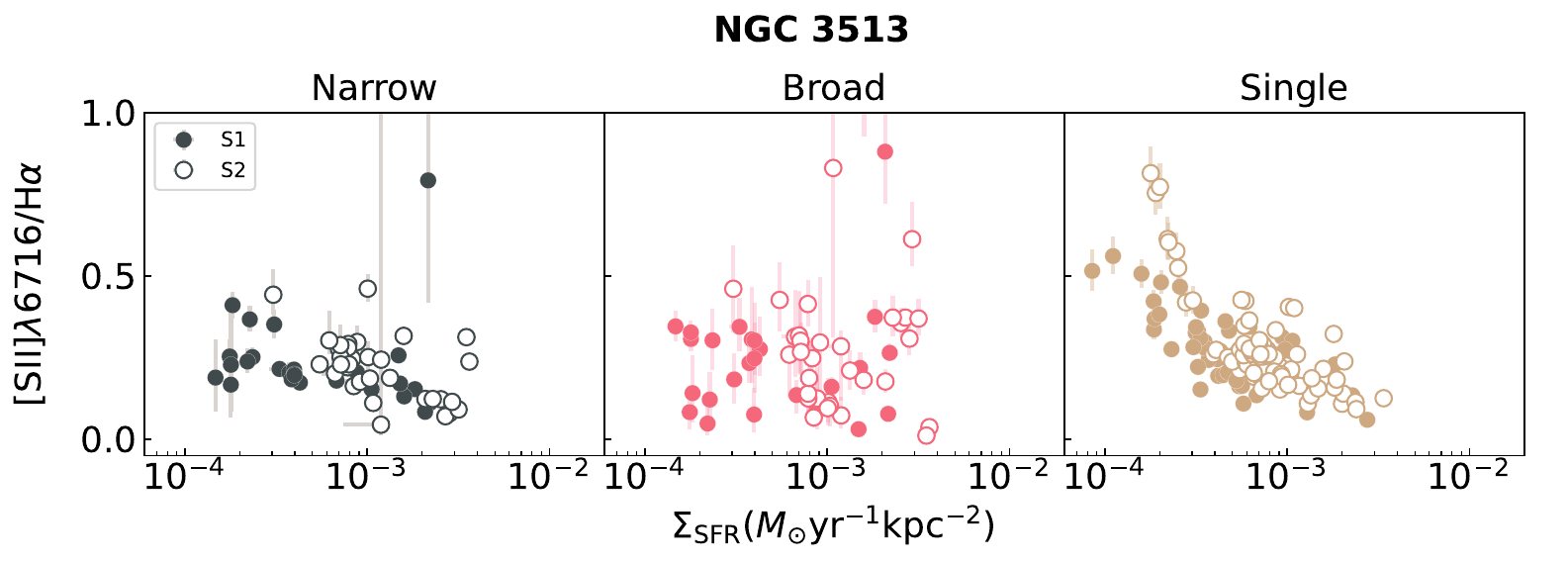} 
\caption{The same as the top panel of Figure ~\ref{fig:recombination_line_compare} but for NGC\,3513. The broad line ratios do not depend on $\Sigma_{\rm{SFR}}$.} \label{fig:line_ratio_SFR_3513}
\end{figure*}

In addition, we display the dependence of [\ion{S}{II}]$\lambda$6716/H$\alpha$ on $\Sigma_{\rm{SFR}}$ in Figure~\ref{fig:line_ratio_SFR_3513}. There is a mild anti-correlation between the emission-line ratios and $\Sigma_{\rm{SFR}}$ in the narrow component, which is stronger in the single component spectra. This is again consistent with observations of planar DIG, where the dilution and reprocessing of the ionizing spectrum away from star-forming regions results in elevated line ratios \citep[e.g.,][]{Haffner2009}. The spatial correlation between the elevated emission-line ratios and the outflow regions discussed above suggests that shocks in the planar DIG may also contribute to the higher [\ion{S}{II}]$\lambda$6716/H$\alpha$ at lower $\Sigma_{\rm{SFR}}$. Finally, we note that there is not a clear trend between the emission-line ratios and the star-formation rate surface density in the broad-component spectra; this correlation is also weak in NGC\,3511, indicating that the local $\Sigma_{\rm{SFR}}$ does not fully determine the physical conditions in the eDIG in either galaxy. This is consistent with our conclusion for NGC\,3511 that the conditions in the eDIG are more significantly shaped by the burstiness of the star-formation history than by the instantaneous star-formation rate.

\section{Constructing Mass Models for NGC\,3511 and NGC\,3513} \label{sec:mass_model}
In this section, we construct mass models for NGC\,3511 and NGC\,3513 to determine their galactic gravitational potentials. We model both the stellar disk component and the dark matter halo component of each galaxy.

We assume that both the vertical ($z$-direction) and radial ($R$-direction) density profiles for the stellar disk are exponential:
\begin{equation}
\rho = \rho_{0}e^{-R/h_{R}}e^{-z/h_{z}}.
\end{equation}
Here $z$ is the distance away from the midplane in the vertical direction, $R$ is the distance from the center in the radial direction, and $\rho_{0}$ is the density of the stellar disk at the disk center ($R = 0$, $z = 0$). $h_z$ is the disk scale height in the vertical direction, and $h_R$ is the disk scale height in the radial direction. For both galaxies, we assume $h_{z}= 0.3$ kpc and $h_{R}=2.4$ kpc, which is consistent with values reported for the Milky Way \citep{BlandHawthorn2016}. We use the stellar mass measurements in Table~\ref{tab:galaxy_properties} to obtain $\rho_{0} = 10^{8.96} \,M_\odot \rm \,kpc^{-3}$ for NGC\,3511 and $\rho_{0} = 10^{8.11} \,M_\odot \rm \,kpc^{-3}$ for NGC\,3513. The corresponding gravitational potential and vertical gravitational acceleration have been derived by \citet{Cuddeford1993}.

For the dark matter halo, we assume a Navarro–Frenk–White (NFW) density profile \citep{Navarro1997}. From the stellar mass-halo mass relation of \citet{Behroozi2019}, we adopt a dark matter halo mass from the stellar mass. We additionally assume that the concentration parameter is tightly correlated with the dark matter halo mass following the \citet{DiemerJoyce19} model. This yields the NFW profile and corresponding gravitational potential and vertical gravitational acceleration. We have confirmed that our parameter choices are reasonable based on comparison with the observed rotational velocity curves.

\bsp
\label{lastpage}

\end{document}